\documentclass[twocolumn,prb,aps,superscriptaddress]{revtex4}

\newcommand{\tis}{TiS$_3$~}
\newcommand{\mos}{MoS$_2$~}
\newcommand{\etal}{\emph{et al.~}}

\newcommand{\bq}{{\bm q}}
\newcommand{\bk}{{\bm k}}

\usepackage{bm}

\usepackage{graphicx}
\usepackage{multirow}
\usepackage{color}
\usepackage{amsmath,amsfonts,amssymb}

\begin{document}

\title{Anisotropic features in the electronic structure of the two-dimensional transition metal trichalcogenide TiS$_3$: electron doping and plasmons}

\author{J. A. Silva-Guill\'{e}n}
\email{josilgui@gmail.com}
\affiliation{Fundaci\'on IMDEA Nanociencia, C/Faraday 9, Campus Cantoblanco, 28049 Madrid, Spain}
\author{E. Canadell}
\affiliation{Institut de Ci\`encia de Materials de Barcelona (ICMAB-CSIC), Campus Bellaterra, 08193 Barcelona, Spain}
\author{P. Ordej\'on}
\affiliation{Catalan Institute of Nanoscience and Nanotechnology (ICN2), CSIC and The Barcelona Institute of Science and Technology, Campus Bellaterra, 08193 Barcelona, Spain}
\author{F. Guinea}
\affiliation{Fundaci\'on IMDEA Nanociencia, C/Faraday 9, Campus Cantoblanco, 28049 Madrid, Spain}
\affiliation{Department of Physics and Astronomy, University of Manchester, Oxford Road, Manchester M13 9PL, UK}
\author{R. Rold\'an}
\email{rroldan@icmm.csic.es}
\affiliation{Instituto de Ciencia de Materiales de Madrid, CSIC, Sor Juana Ines de la Cruz 3, 28049 Cantoblanco, Madrid, Spain}

\begin{abstract}
Analysis of the band structure of TiS$_3$ single-layers suggests the possibility of changing their physical behaviour by injecting electron carriers. The anisotropy of the valence and conduction bands is explained in terms of their complex orbital composition. The nature of the Fermi surface and Lindhard response function for different doping concentrations is studied by means of first-principles DFT calculations. It is suggested that for electron doping levels $x$ (number of electrons per unit cell) $\sim$ 0.18-0.30$e^-$ the system could exhibit incommensurate charge or spin modulations which, however, would keep the metallic state whereas systems doped with smaller $x$ would be 2D metals without any electronic instability. The effect of spin-orbit coupling in the band dispersion is analysed. The DFT effective masses are used to study the plasmon spectrum from an effective low energy model. We find that this material supports highly anisotropic plasmons, with opposite anisotropy for the electron and hole bands.

\end{abstract}

\maketitle

\section{Introduction}\label{introduction}

\begin{figure}
\centering
\includegraphics[scale=0.5]{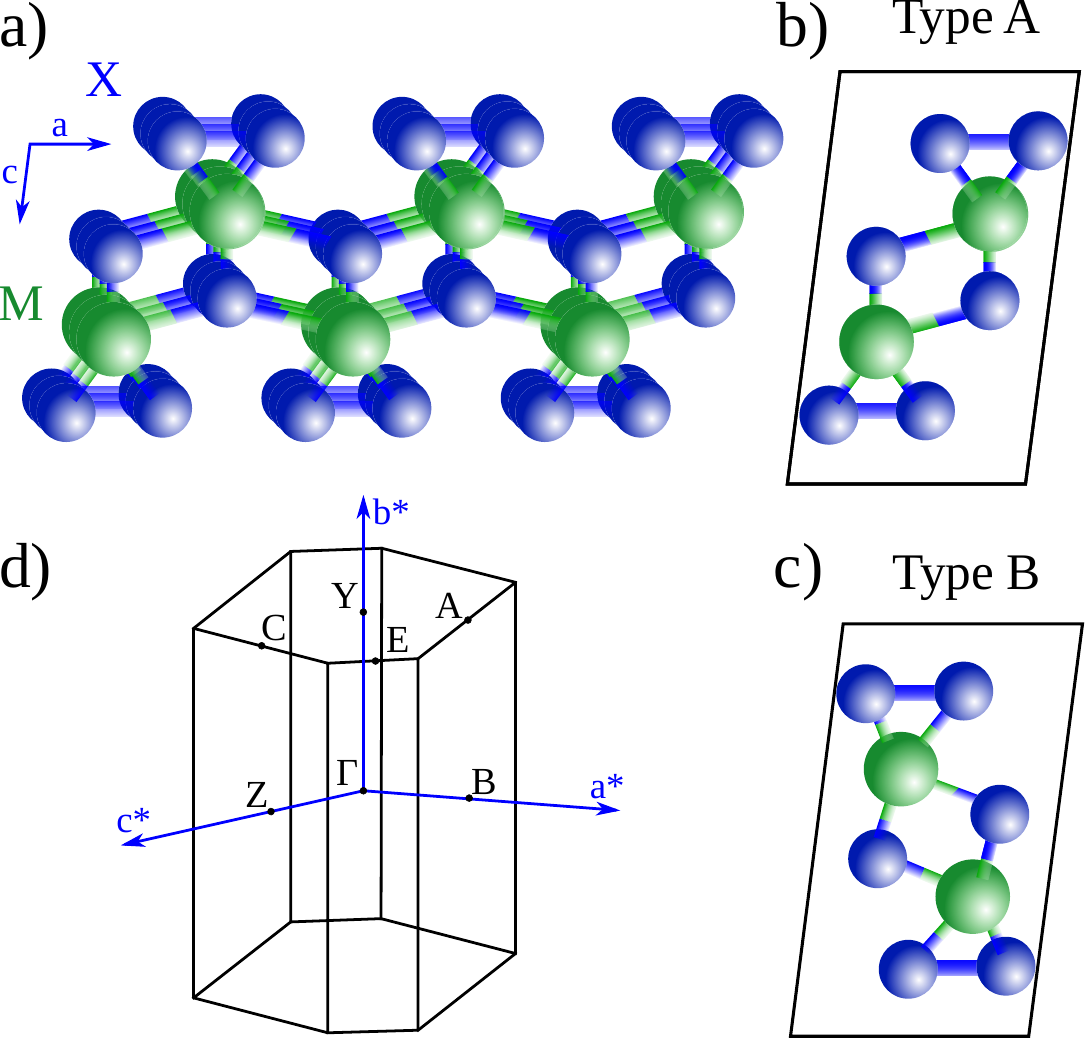}
\caption{Crystal structure of bulk group IVB TMTCs MX$_3$ (a) showing the two proposed structure types A (b) and B (c). The transition metal atoms are in a bicapped trigonal prismatic environment which is more symmetric in the type A structure. (d) Schematic Brillouin zone of the system.}
\label{fig:struct_A_B}
\end{figure}

Since the discovery of graphene in 2004,\cite{review_graphene_2009} there has been a huge improvement in the fabrication and manipulation of layered materials.\cite{geim_SC_2015} Recently, the discovery that a monolayer of \mos changes its electronic properties with respect to the bulk\cite{mak_mos2_2010} brought much expectation in the scientific community towards the transition metal chalcogenides (TMCs). The MX$_2$ transition metal dichalcogenides (TMDCs) have been thoroughly studied both experimentally\cite{geim_SC_2015,Geim_PNAS_2005,Review_Wang_2012,mak_mos2_2010} and theoretically.\cite{roldan_review_2014,Guo_2014,Zhou_2012,Rosner_2014} Nowadays, TMCs with different chemical stoichiometries such as the transition metal trichalcogenides (TMTCs) MX$_3$ (Fig. \ref{fig:struct_A_B}) are also being intensely studied. \cite{dai_2016,Island2017Electronics} Interestingly, \tis \cite{TiS3_Castellanos_2015} has shown to have cleavage energies close to that of graphite, showing that similar methods can be used to fabricate \tis monolayers.\cite{jin_2015}

The electronic structure of bulk group IVB TMTCs, MX$_3$ (M= Ti, Zr, Hf; X= S, Se, Te), has been the subject of several theoretical studies\cite{Bullett_1979,Myron_1981,Canadell_1988,Canadell_1989,Stowe_1998,Felser_1998} but those concerning slabs of different thicknesses are more scarce. \cite{jin_2015,dai_2015,TiS3_Castellanos_2015,Li_Nano_2015,Peeters_2015,Kang_Jun_2016} These works have mostly dealt with the variation of the band gap in single-layers when changing the nature of both the transition metal and chalcogenide atoms or applying strain, the mechanical properties, and the role of vacancies. Some interesting tendencies have already been pointed out.\cite{dai_2016} For instance, among the semiconducting members of this family (i) the single-layer band gap diminishes when the size of the chalcogenide increases, (ii) the bulk indirect band gap may change to direct in the single-layers (TiS$_3$) or keep its indirect nature (TiSe$_3$), and (iii) the indirect band gap of ZrS$_3$ and HfS$_3$ can undergo an indirect-to-direct band gap transition with increasing tensile strain whereas the corresponding selenides keep their indirect band gap.

From a more fundamental viewpoint the potential of slabs with a few layers of group IVB TMTCs is still far from being fully explored. The conduction band of the semiconducting systems originates from the $d$ orbitals and, not surprisingly,  the carrier mobility associated with electrons in the pristine TiS$_3$ single-layer seems to be very anisotropic, with better mobility along the trigonal prismatic chains.\cite{jin_2015} Therefore, it is expected that under electron doping the system may become an anisotropic conductor and thus be subject to the electronic instabilities of pseudo-one-dimensional (pseudo-1D) conductors. If this is the case, it would be interesting to see how much 1D would be these doped single-layers, i.e. would they exhibit metal to insulator or metal to metal transitions? It may be expected that the trigonal prismatic chains become more strongly coupled when the size of the chalcogen atom increases. The doped system could be then less 1D so that the physical behaviour of the doped single-layers could be modulated by varying the chalcogen atoms. At this point, let us remind that the Zr and Hf tellurium compounds are already metallic even without doping.\cite{Sambongi_1986} In fact, bulk ZrTe$_3$ exhibits a charge density wave (CDW)\cite{Sambongi_1984} at 63 K (whose origin is still debated) and bulk superconductivity under pressure ($T_c$ = 4 K at 5 GPa).\cite{Yomo_2005}  

 The family of TMTCs with the TiS$_3$ structure may thus provide a challenging series of compounds whose physical behavior under the form of single-layers or slabs with few layers can provide interesting and new phenomena besides the tunable optical properties. In the present work we explore such possibilities by means of first-principles density functional theory (DFT) calculations for TiS$_3$ single-layers with different doping degrees. We start with a detailed analysis of the electronic band structure, in which the origin of the band anisotropy is discussed in terms of the orbital character of the valence and conduction bands. The results are used to study the screening properties and possible sources of instabilities of doped samples. We finally analyse the nature of the plasmon modes for $n$ and $p$ doped samples, finding a strong anisotropy in the spectrum, which is opposite for the valence and the conduction band modes.   
 
\section{Computational details}\label{comp_meth}

First principles calculations were carried out using a numerical atomic orbitals approach to DFT,~\cite{HohKoh1964,KohSha1965} which was developed for efficient calculations in large systems and implemented in the \textsc{Siesta} code.\cite{SolArt2002,ArtAng2008} 
We have used the generalized gradient approximation (GGA) and, in particular, the functional of Perdew, Burke and Ernzerhof.~\cite{PBE96} 
Only the valence electrons are considered in the calculation, with the core being replaced by norm-conserving scalar relativistic pseudopotentials~\cite{tro91} factorized in the Kleinman-Bylander form.\cite{klby82} The non-linear core-valence exchange-correlation scheme~\cite{LFC82} was used for all elements. We have used a split-valence double-$\zeta $ basis set including polarization functions.\cite{arsan99} The energy cutoff of the real space integration mesh was set to 500 Ry. To build the charge density (and, from this, obtain the DFT total energy and atomic forces), the Brillouin zone (BZ) for the bulk and monolayer was sampled with the Monkhorst-Pack scheme\cite{MonPac76} using grids of (30$\times$30$\times$30) and (30$\times$30$\times$1) {\it k}-points, respectively.  

The Lindhard response function:
\begin{equation}\label{Eq:Chi}
\chi({\bq})=-\sum_{i,j}\sum_{\bk}\frac{f_F(\epsilon_i({\bk}))-f_F(\epsilon_j({\bk}+{\bq}))}{\epsilon_i({\bk})-\epsilon_j({\bk}+{\bq})},
\end{equation}
where $f_F$ is the Fermi function, was obtained from the computed DFT values of the band eigenvalues $\epsilon_i({\bk})$. The integral over {\it k}-points of the BZ was approximated by a direct summation over a regular grid of points. As the Lindhard function is more sensitive to the accuracy of the BZ integration than the total energy, especially in very anisotropic systems, and/or in the presence of hot spots in the band structure (e.g. saddle points with the corresponding van Hove singularity in the DOS), the {\it k}-points grid used for its calculation must be more dense. The calculations are done, nevertheless, using the eigenvalues obtained in the DFT calculation for the coarser grid, and interpolating their values in the denser grid, using a post-processing utility available within the \textsc{Siesta} package. In this work, for the calculation of the Lindhard function of the single layers, the BZ  was sampled using a grid of (90$\times$90) {\it k}-points in the layer plane. The two lower conduction bands, which are those becoming partially filled for the electron doping levels considered in this work, were those taken into account in the calculations. 

For the structural relaxations, we maintain the known  symmetry of the crystal structure. For the bulk structures, as the weak interaction between layers is known to be severely underestimated by the GGA functionals, we have maintained the experimental value for the $c$ lattice parameter, while allowing the in-plane lattice parameters and the internal atomic coordinates to vary.

We also performed calculations using the HSE06\cite{HSE06} functional implemented in the Vienna ab initio simulation package (VASP)\cite{vasp} in order to obtain the corrected values of the gap. The plane wave cutoff was set to 340 eV. The Brillouin zone for these calculations was sampled using a grid of ($7\times10\times4$) and ($7\times10\times1$) {\it k}-points for the bulk and monolayer structures, respectively.

\begin{figure*}
\includegraphics[scale=0.9]{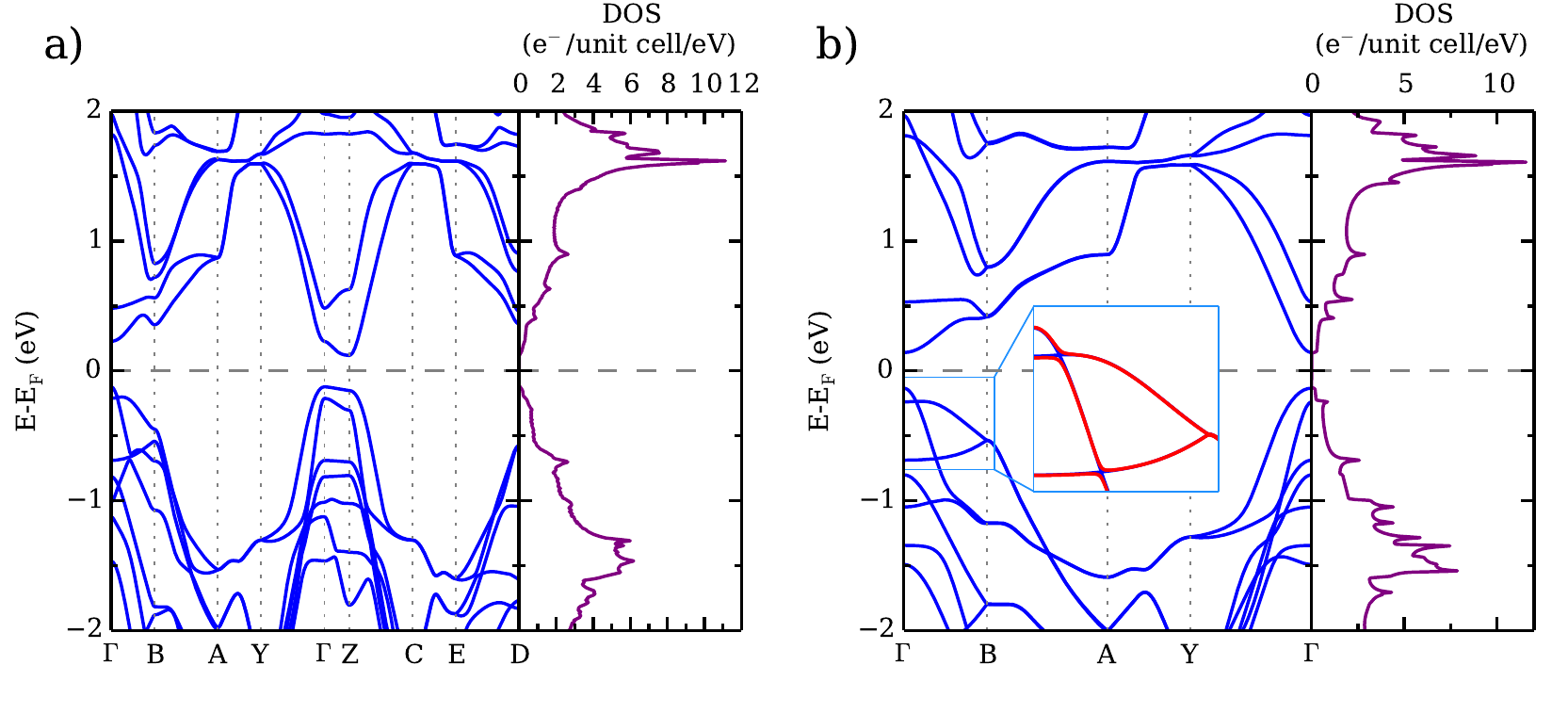}
\caption{Calculated band structures of the bulk (a) and single-layer (b) structures of TiS$_3$. $\Gamma$ = (0, 0, 0), B = (1/2, 0, 0), A = (1/2, 1/2, 0), Y = (0, 1/2, 0), Z = (0, 0, 1/2), C = (0, 1/2, 1/2), E = (1/2, 1/2, 1/2) and D = (1/2, 0, 1/2) in units of the reciprocal lattice vectors and the Fermi level is set at zero. The inset in (b) shows the only noticeable changes in the band structure when the SOC is taken into account. The bands calculated with/without SOC coupling are shown in red/blue, respectively.}
\label{fig:vector_projection}
\end{figure*} 

\section{Results and discussion}

\subsection{Structural aspects}

The MX$_3$ layers of group IVB TMTCs (MX$_3$) are built up from trigonal prismatic MX$_3$ chains in such a way that two rectangular faces of an MX$_6$ trigonal prism are capped by X atoms of the neighboring chains (see Fig. \ref{fig:struct_A_B}). Thus, every transition metal atom is coordinated to eight chalcogen atoms, X. One of the three X-X distances of the triangles is short and compatible with a single bond X-X so that for electron counting purposes this X-X pair must be considered as (X-X)$^{2-}$. The system can thus be formulated as M$^{4+}$(X-X)$^{2-}$X$^{2-}$ and should be a semiconductor. Furuseth and coworkers\cite{furuseth_75} found that these MX$_3$ systems may have two structural variations, named type A and type B (see Fig. \ref{fig:struct_A_B}). In the type A structure the triangles and the two capping M-X distances are practically symmetric with respect to a plane containing the M and ``isolated'' X atoms of a chain. This plane is not really a symmetry plane of the structure because of the monoclinic crystal structure. In the type B structure any trace of this symmetrical situation is lost. Some compounds of this family were found to exhibit the type A structure (ZrSe$_3$, ZrS$_3$) whereas others the type B structure (TiS$_3$, ZrTe$_3$).\cite{furuseth_75} However, later work concerning one type B structure (ZrTe$_3$) showed that the structure was really of type A, casting doubts about the existence of two different structure types. \cite{furuseth_91} The inter and intrachain S-S contacts are considerably different in the two structures, so that the structural type can influence the location of the top of the valence band which is mostly chalcogen-based. Hence, we first decided to revisit the structure of bulk TiS$_3$ by looking in detail to this point. Geometry optimizations with different starting geometries including those of the type B structure led always to a type A structure with geometrical parameters similar to those reported by Jin \etal\cite{jin_2015} Thus, we confirm that TiS$_3$ does not exhibit type B structure but the more symmetric type A and this is most likely the case for all TMTCs of this group. The calculations reported in the following sections use the bulk and single-layer DFT optimised structure.

\begin{table}
    \centering
     \caption{Effective masses for the top of the valence band and bottom of the conduction band of the TiS$_3$ single-layer in units of the electron rest mass ($m_{e^-}$).}
    \begin{tabular} {lccccc}
    \hline
    \hline
    & \hspace*{0,5cm} &    V.B.      & \hspace*{0,5cm} &      C.B.    & \hspace*{0,3cm} \\
    \hline
    $\Gamma$-B   &\hspace*{0,5cm} &   0.3317 &\hspace*{0,5cm} &   1.8718 \\
    Y-$\Gamma$   &\hspace*{0,5cm} &   0.8333 &\hspace*{0,5cm} &   0.5253 \\
    \hline
    \hline
    \end{tabular}
   
    \label{tab:my_label}
\end{table}

\begin{figure*}
\centering
\includegraphics[scale=0.85]{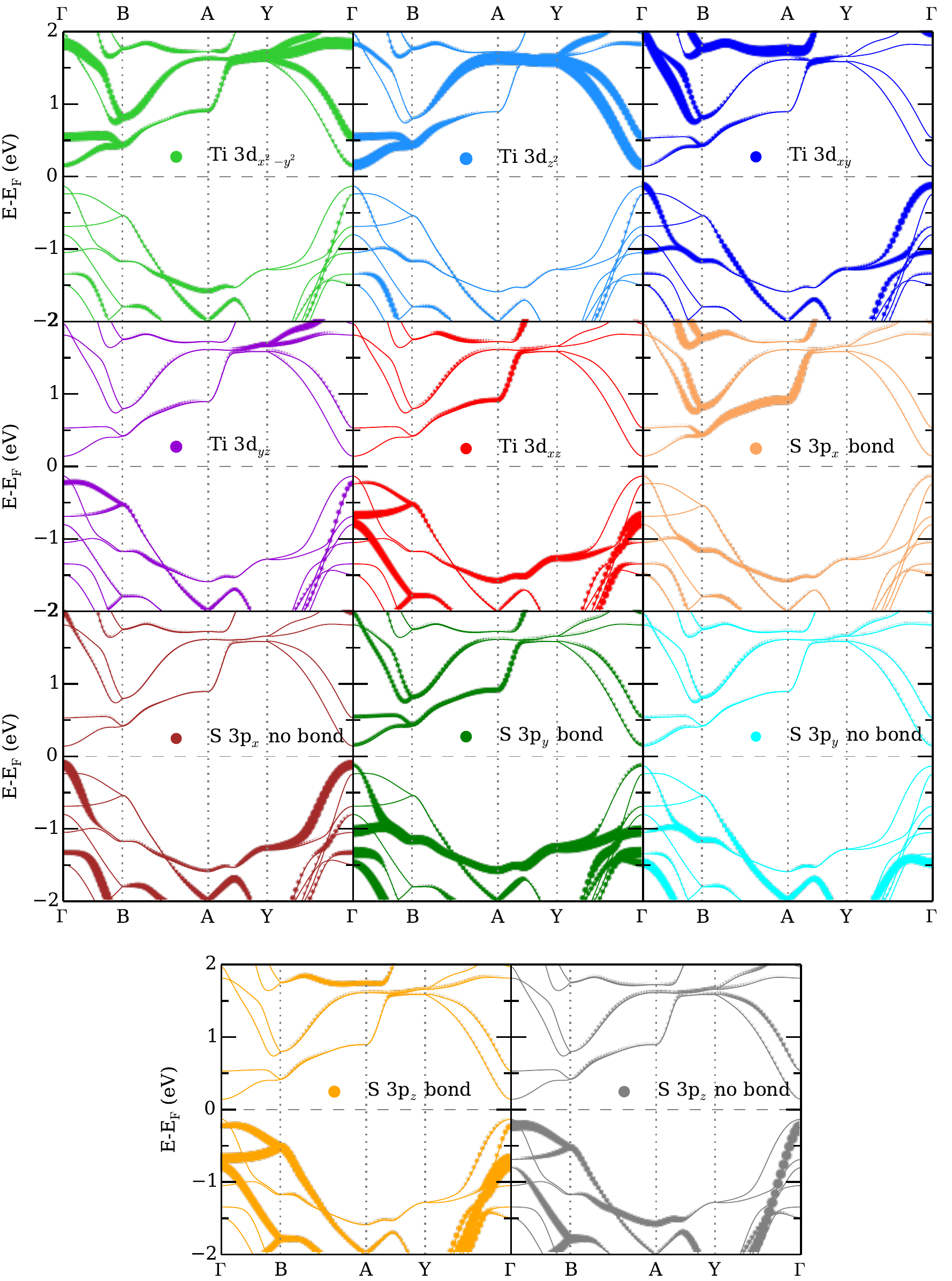}
\caption{Fatbands analysis of the single-layer band structure. $\Gamma$ = (0, 0, 0), B = (1/2, 0, 0), A = (1/2, 1/2, 0) and Y = (0, 1/2, 0) in units of the reciprocal lattice vectors.}\label{fig:mono_FB}
\end{figure*}

\subsection{Nature of valence and conduction bands}\label{Sec:Nature}

Shown in Fig.~\ref{fig:vector_projection} are the calculated band structures and density of states (DOS) for the optimized structures of TiS$_3$ bulk (a) and single-layer (b). In our calculations we find an indirect gap of 225 meV from $\Gamma$ to Z for the bulk, and 
a direct gap of 246 meV at $\Gamma$ for the single layer. The calculated effective masses of the top of the valence band and bottom of the conduction band for the TiS$_3$ single-layer are shown in Table~\ref{tab:my_label}. We find that the mass anisotropy ratio, obtained from the effective masses along the $a$ and $b$ crystallographic directions $m_a/m_b$, is opposite for the valence and conduction bands, with respective values of $\sim 0.4$ and $\sim 3.56$. All these results are in good agreement with previous plane wave-based DFT calculations.\cite{dai_2015,Li_Nano_2015,Peeters_2015,Kang_Jun_2016} We have also checked that inclusion of spin-orbit coupling (SOC) does not lead to any substantial variation, as it can be seen in Fig.~\ref{fig:vector_projection}b (inset), where we compare the electronic band structure in the presence (red) and in the absence (blue) of SOC. Nevertheless it is important to notice that SOC leads to several band avoided crossings. This is specially important for the two branches of the valence band along the $\Gamma$-B direction, for which SOC leads to a sizeable splitting of $\sim18$ meV.  

Because of the well known underestimation of these band gaps when using the PBE functional\cite{dai_2015} we have recalculated the bulk and single-layer band structure using the  HSE06 functional.\cite{HSE06} The calculated gaps are 1.09 and 1.12 eV for the bulk and monolayer structures, respectively. These values are in good agreement with the experimental bulk value, $\sim$~0.9~eV.\cite{opt_gap_1961} The band structures calculated with the two functionals are practically identical except for an almost rigid shift of the empty bands so that from now on our analysis is based on the PBE type calculations.

\begin{figure}
\centering
\includegraphics[scale=0.9]{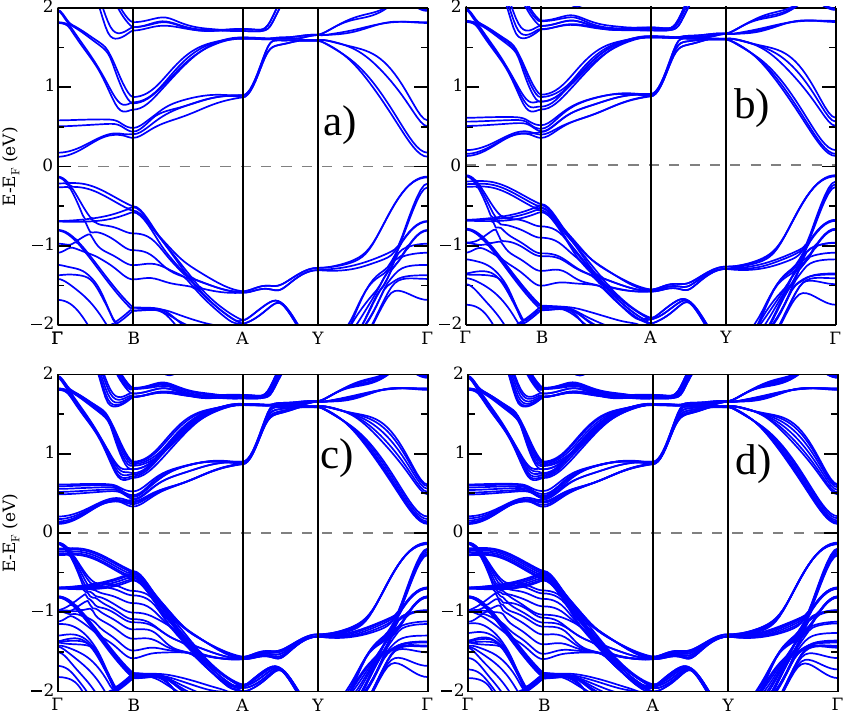}
\caption{Calculated band structure for slabs with two (a), three (b), five (c) and six (d) TiS$_3$ layers.}\label{fig:Multilayer}
\end{figure}

\begin{figure*}
\includegraphics[scale=0.9]{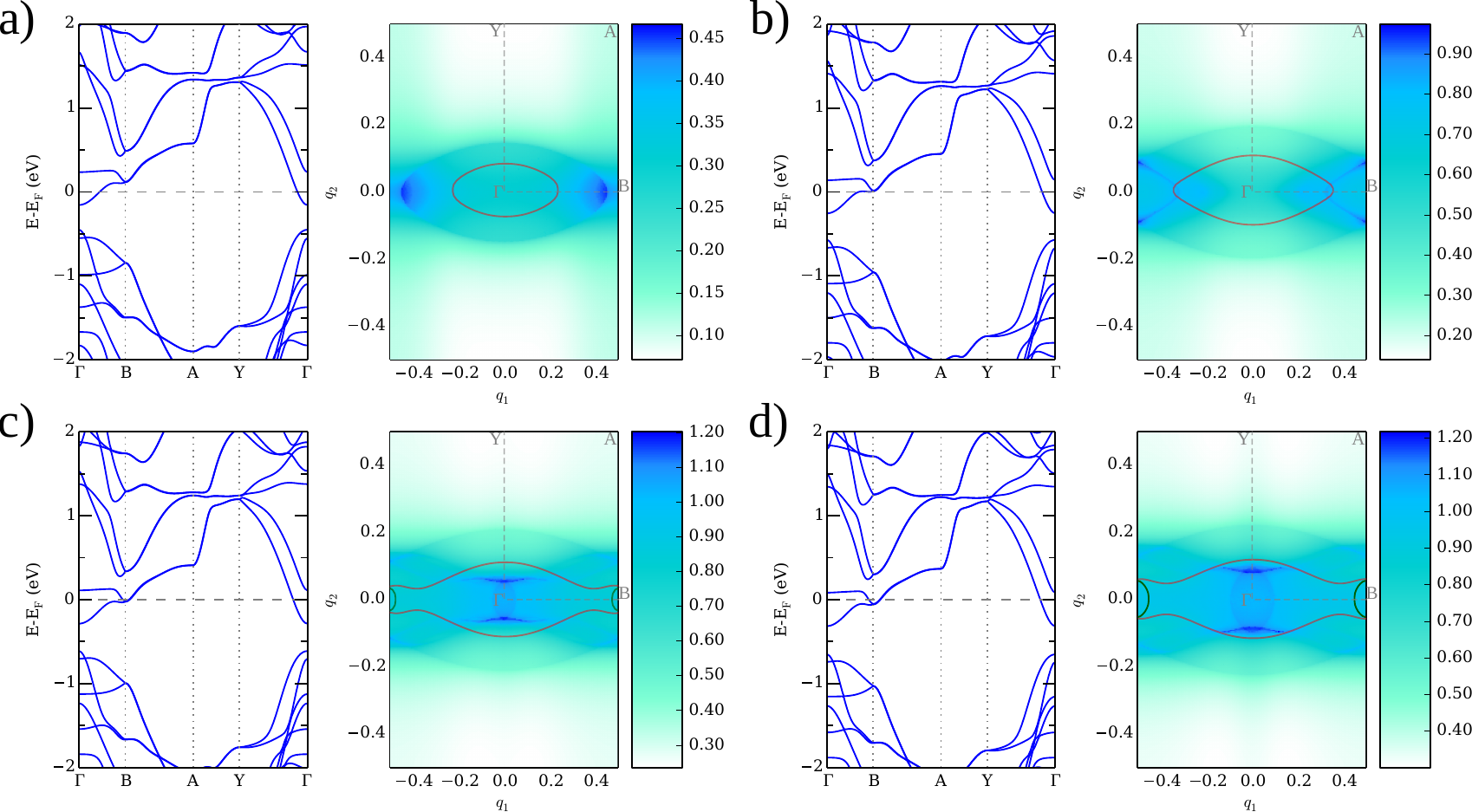} 
\caption{TiS$_3$ single-layer doped with $+0.1$ (a), $+0.2$ (b), $+0.25$ (c) and $+0.3$ (d) $e^-$ per unit cell. Left: Band structure considering the re-optimization of atomic positions for each $x$ value; Right: Lindhard function with the Fermi surface (continuous red and green lines) superposed. The horizontal and vertical axes are the $a$*- and $b$*-directions.}
\label{fig:doping_0_1}
\end{figure*}

 It is important for our discussion to clearly grasp the relationship between the nature of the valence and conduction bands and the structural parameters. Shown in Fig. \ref{fig:mono_FB} is a fatband analysis of the single-layer band structure. The orbitals are specified according to a local system of axes such that the x-axis coincides with the $a$-direction and the z-axis goes along the direction of the chains, i.e., the $b$-direction. The top of the valence band is mostly made of the S 3$p_x$ orbitals and a smaller contribution of the Ti 3$d_{xy}$ orbitals. In addition, the S 3$p_x$ contribution comes only from the inner sulfur atoms (i.e. those not forming S-S bonds; see Fig. \ref{fig:mono_FB}). Thus, the valence band concentrates in the inner part of the layer and is mostly directed along the $a$-direction. Consequently, the top of the valence band should exhibit very weak interlayer interactions when the number of layers increases. This is indeed what we obtain in our DFT calculations for multilayer samples, shown in Fig. \ref{fig:Multilayer}. Notice that the edge of the upper valence band appears basically degenerated for multilayer samples, whereas the second valence band suffers a noticeable splitting into well separated subbands, the number of which depends on the number of stacked layers. Since the main orbital contribution to this band comes from S 3$p_z$ of both inner and outer sulphur atoms, which play an essential role in inter-layer hopping, this leads to a splitting of this band in multilayer samples. In addition, the effective mass of the hole carriers should be considerably smaller along the $a$- than the $b$-direction of the layer, as shown in Table \ref{tab:my_label}. The conduction band is strongly based on the Ti 3$d_{z^2}$ orbitals which point along the $b$-direction and this confers a strong anisotropy to the electron carriers. However, the conduction bands have also an important participation of the Ti 3$d_{x^2-y^2}$ and to a lesser extent from the S 3$p_y$ orbitals of the sulphur atoms of the outer part of the layer (i.e. those implicated in the S-S bonds; see Fig. \ref{fig:mono_FB}). Such S 3$p_y$ contribution is the responsible for the downward shift of the valence band along the $\Gamma $-Z direction in the bulk and it further leads to splitting of the conduction band edges in multilayer samples, as it can be seen in Fig. \ref{fig:Multilayer}. Of course, the effective mass of the electron carriers is anisotropic and smaller along the $b$-direction (see Table \ref{tab:my_label}).

\subsection{Electron doping}

The fatband analysis of the TiS$_3$ band structure is challenging in that it highlights the possibility of altering the physical behavior of TiS$_3$ single-layers by injecting electron carriers in the conduction band through electric field gating. The dominance of Ti 3$d_{z^2}$ orbitals in the conduction band points out toward a strongly anisotropic metallic behavior (i.e. better conductivity along the $b$-axis) with associated inherent electronic instabilities under such conditions. However, one must bear in mind that the number of carriers which can be injected by gating has some limitations. Assuming, for instance, doping levels ($x$) similar to those attained in MoS$_2$\cite{MoS2_doping} it would lead to $x$ values of up to $\sim$ 0.3$e^-$ per unit cell (i.e. per two TiS$_3$ units). This would result with Fermi levels occurring between the bottom of the conduction band and the energy where the two lower conduction bands become degenerate at the B point (see Fig. \ref{fig:vector_projection}b). In view of the non-negligible dispersion of the lower conduction band along the $a^*$-direction, at least for low and intermediate doping levels which will not lead to open Fermi surfaces, it is not clear what will be the detailed topology of the Fermi surface. In order to explore this issue we have calculated the Fermi surface and Lindhard response function for different doping levels. The atomic positions were re-optimized for every value of $x$ although it is not expected that structural effects can play an important role for these low doping levels.  

The calculated band structure, Fermi surfaces and Lindhard response function for several electron doped TiS$_3$ single-layer systems with doping levels $x$ = 0.1 to 0.3$e^-$ are shown in Figs. \ref{fig:doping_0_1}a-d. In these figures we superpose the calculated Fermi surface and the Lindhard response function. The portions of the Fermi surface originating from the lowest and second lowest conduction bands are shown in red and green, respectively. As can be seen from the almost identical calculated band structures, electron doping with these $x$ values practically does not alter the band structure. If the conduction band was a perfect 1D system the Fermi surface would consist of two parallel lines perpendicular to the $b$*-direction at $\pm$ 0.025$b$* ($x$ = 0.1$e^-$) and 0.05$b$* ($x$ = 0.2$e^-$). However, this is very far from the computed results. From $x$ = 0 to $x$ = 0.1$e^-$ the Fermi surface is closed (Fig. \ref{fig:doping_0_1}a), meaning that there are non-negligible inter-chain interactions. The Fermi surface is an ellipse with the long axis along $a$* essentially because the slope of the band is larger along $b$*. The Lindhard function has a somewhat broad region with larger values for wave vectors corresponding to the nesting of the locally flat borders of the ellipse at 0.22$a$*. However, these values are relatively small and it is not expected that they will lead to any instability. Thus, for this doping interval the system should be a 2D metal (except for very small values of $x$ where the system may exhibit activated conductivity because of the low density of carriers and the potential due to random sulfur vacancies). 

When $x$ increases, the Fermi surface is still closed but clear 1D features already appear. The  shape  of  the  ellipse  changes  in  such  a  way  that  long  flat  portions  occur (see Fig. \ref{fig:doping_0_1}b for $x$ = 0.2$e^-$). This is due to the fact that the density of carriers has increased and the Fermi level is
now reaching a region where the slope along the chains direction ($\Gamma$-Y) strongly dominates over the slope along the inter-chain direction ($\Gamma$-B), i.e. a region where the electron gas already exhibits a pseudo-1D behaviour. However, the Fermi surface is not made of two cosine-like lines as it could be expected for such a case. This is due to the fact that the coupling between chains is large enough to outweigh the effect of the carrier increase and the Fermi surface is ultimately closed even if the pseudo-1D character inherent to the valence band already shows up. The flat sections are well nested and lead to maxima of the Lindhard function which for $x$ = 0.2$e^-$ occur at $\pm$ 0.5$a$*$\pm $0.085$b$* (see Fig. \ref{fig:doping_0_1}b). Thus, for values of $x$ approaching 0.2$e^-$, electronic instabilities partially destroying the Fermi surface originating from these nesting vectors can occur.

Following the previous reasoning one could expect that for larger values of $x$ the pseudo-1D character of the system would clearly appear as an open and well nested Fermi surface made of two cosine-like lines. Slightly above 0.20$e^-$ the Fermi surface indeed becomes open but the shape is not the expected one. That corresponding to 0.25$e^-$ is shown in Fig. \ref{fig:doping_0_1}c. Note that for doping levels just above $x$ = 0.22$e^-$ the Fermi level cuts also the second conduction band around the B point. For very low band fillings this band exhibits similar curvature along the B-$\Gamma$ and the B-A directions so that a closed component around the B point emerges (the green lines of the Fermi surface). The shape of the portion originating from the first band strongly reminds that of Fig. \ref{fig:doping_0_1}b but around $\pm$~0.38$a$* exhibits two minima. In fact, there are these regions between $\pm$ 0.38$a$* and B which are responsible for the maxima of the Lindhard function, which exhibits two arcs with maxima in the center, at $\pm $0.05$b$. The Fermi surface keeps this general shape until $x$ values of 0.3$e^-$ (Fig.  \ref{fig:doping_0_1}d) and above. Except for the increase in the separation of the warped lines, the only difference is that when the area of the closed portion around B increases, the circle becomes an ellipse. The Lindhard function of systems with doping levels between 0.22$e^-$ and 0.3$e^-$ is the same although the separation of the arcs increases with the doping (they occur at $\pm $ 0.02$b$* for $x$ = 0.22$e^-$ but $\pm $0.085$b$* for $x$ = 0.3$e^-$). This nesting vector occurs in the direction of the trigonal prismatic chains, $b$*, but the modulus is smaller than it would be if the system was really a pseudo-1D system (even if taking into account the contribution of the small closed part around B; see Section \ref{coupling}). These results suggest that for values of $x$ between 0.22 and 0.3$e^-$ there could occur an instability associated with this nesting vector which is very different from that discussed for lower carrier concentrations. This nesting vector would be responsible for the destruction of the flat portion of the Fermi surface around the border of the Brillouin zone. Note than when the closed portion of the Fermi surface around B appears, the Lindhard function has an additional contribution around the center of the arcs due to partial nesting of this closed part (for similar reasons as those discussed for Fig. \ref{fig:doping_0_1}a). To summarize, we believe that systems doped with $x$ values around 0.18-0.2 $e^-$ and between 0.22-0.3$e^-$ could exhibit two different types of incommensurate modulations of the charge density wave (CDW) or spin density wave (SDW) type which, however, would keep the metallic state whereas systems doped with less than around 0.18-0.2 $e^-$ would be 2D metals without any electronic instability. 

\begin{figure*}
\centering
  \includegraphics[scale=0.9]{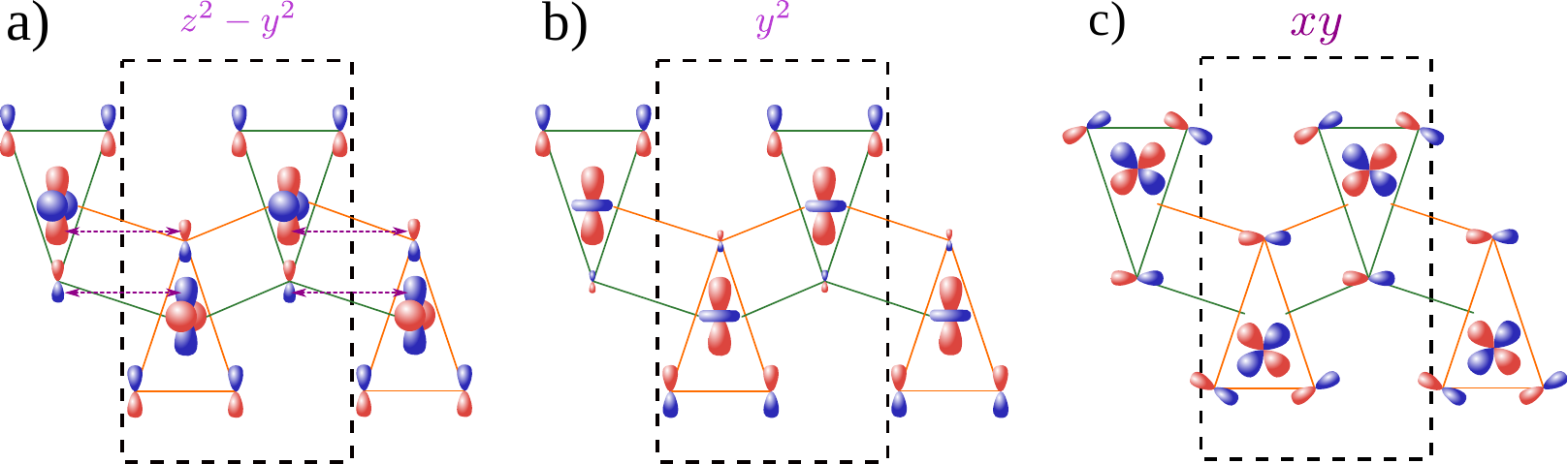} 
\caption{Schematic representation of the lowest (a), second lowest (b) crystal orbitals at $\Gamma $, and the strongly descending Ti 3$d_{xy}$ based band at $\Gamma $ (c). The triangles shown in green and orange are displaced by $b$/2 so that all atom-atom distances with the same color occur in the same plane.}
\label{fig:orbitals}
\end{figure*}

\subsection{Inter-chain coupling and the conduction band}\label{coupling}
 
As discussed above, the Fermi surfaces for the doping levels considered here are different from those intuitively expected meaning that the interchain interaction is not a trivial one. As shown in Fig. \ref{fig:vector_projection}b, the two lowest empty bands exhibit the typical behavior along $b$* for a highly dispersive band based on the Ti 3$d_{z^2}$ orbitals. The separation of the two bands at $\Gamma$ is around 0.3 eV which is non negligible but not exceptionally strong. Looking at the $\Gamma$--B direction (see Fig. \ref{fig:mono_FB}) one notices that the two lowest bands essentially result from a strong mixing of the 3$d_{z^2}$ and  3$d_{x^2-y^2}$ orbitals, leading to a moderately dispersive and a dispersionless pair of bands, respectively. However, in the close vicinity of B there is an avoided crossing with a strongly dispersive 3$d_{xy}$ based pair of bands. Thus, it appears that the nature of the lowest conduction bands along the interchain direction ($\Gamma$--B) changes significantly, acquiring 3$d_{xy}$ character in the last part of the line and this is at the origin of the unexpected behavior.

As it is well known,\cite{NbSe3,Canadell_1989} the fact that the sulfur triangles are far from equilateral, does not only change the electron counting but also induces an important rehybridization of two of the three low-lying levels of the transition metal, 3$d_{z^2}$ and  3$d_{x^2-y^2}$. The reason is that these levels try to minimize the antibonding interactions with the S 3$p$ levels. When the Ti atoms become eight coordinated because of the two additional Ti-S bonds between chains, one of these two levels becomes well separated from the other Ti-based levels and spreads out to lead to the two lowest bands of the undoped system. Depending on the degree of mixing between the two orbitals, the resulting one can be described as 3$d_{z^2-y^2}$ (for moderate mixing) or 3$d_{y^2}$ (for a strong mixing). Analysis of the fatbands and the detailed crystal orbitals clarify the shape of the band structure in that zone. The two lowest empty crystal orbitals of the undoped system at $\Gamma $ are schematically shown in Fig. \ref{fig:orbitals}a and b. The lowest level is based on an out-of-phase combination of 3$d_{z^2-y^2}$ orbitals. These orbitals make moderately stabilizing lateral interactions with the 3$p_y$ orbital of the non-bonded S atoms of the two adjacent cells, which are at the same level (the interactions marked in orange in Fig. \ref{fig:orbitals}). Half of these interactions occur within the cell and half occur between cells. When the phase changes from $\Gamma $ to B, the intercell interactions become antibonding, so that globally this band is non bonding between chains and consequently the band goes up in energy. 
In principle, the interaction between the 3$p_y$ orbital of the inner S atom of one chain and those of the bonded outer S atoms of the two neighboring chains could also contribute to the energy raising since the intracell overlap is positive both in $\Gamma$ and B but the intercell one changes to antibonding in B. However, the distance between the two sulfur atoms is large because the S atoms are located in planes differing by half-$b$  and the effect is small. In fact, there is a shorter  S--S interchain  contact between the bonded S atoms in second-neighboring chains along $a$. This $\pi$-type interaction could also contribute to the energy raising since they are in-phase at $\Gamma$ but out-of-phase at B. However, this contribution turns out to be extremely weak as shown by the fact that it also occurs for the next band (see Fig. \ref{fig:orbitals}b) which is, nevertheless, very flat along $\Gamma$-B.
Thus, for around two thirds of the $\Gamma $ to B traject, the lowest conduction band behaves as resulting from a series of slightly coupled strongly 1D systems located in the trigonal prismatic chains and the closed nature of the Fermi surface is simply due to the joint effect of the low density of carriers and the non-negligible interaction between chains of 3$d_{z^2-y^2}$ orbitals. 

The next empty crystal orbital is based on an in-phase combination of 3$d_{y^2}$ orbitals. An important difference with the previous band is that now the 3$p_y$ orbital of the non-bonded S atoms practically does not mix into this crystal orbital so that the different chains become uncoupled and hence, the completely flat nature of this band. Finally, near the B point a strongly descending pair of bands based on the Ti 3$d_{xy}$ orbitals (see Fig. \ref{fig:orbitals}c), which make strong interactions with the 3$p_x$ orbitals of the adjacent non bonded S atoms near $\Gamma $ but not around B, undergoes an avoided crossing with the lower pair of bands. As a result, Ti 3$d_{xy}$ character is built up in the two lower bands near the region of the B point stabilizing the two lower bands.  More importantly, the antibonding interaction between the bonded S atoms of two successive chains along $a$ at $\Gamma $ (see  Fig. \ref{fig:orbitals}c) becomes bonding around B. Thus, the avoided crossing brings about an effective decrease of the inter-chain coupling in the region around B. The abnormal shape of these bands in this region is thus the result of the competition between destabilizing interactions between the Ti and capping S atoms (see Fig. \ref{fig:orbitals}a) and stabilizing direct S--S interactions of two S--S bonded pairs in adjacent chains along the interlayer direction. 
It is important to note that the slope of the strongly descending pair of bands will increase when the chalcogen atom is bulkier, i.e. for TiSe$_3$ vs. TiS$_3$, or under some contraction along the $a$ axis. In these cases the bottom of the conduction band will most likely occur around B. This will have a very strong influence both on the Fermi surface of the electron doped system and the optical properties.

\subsection{Plasmons in Single-Layer ${\rm {\bf TiS_3}}$}

\begin{figure}
\centering
\includegraphics[width=80mm]{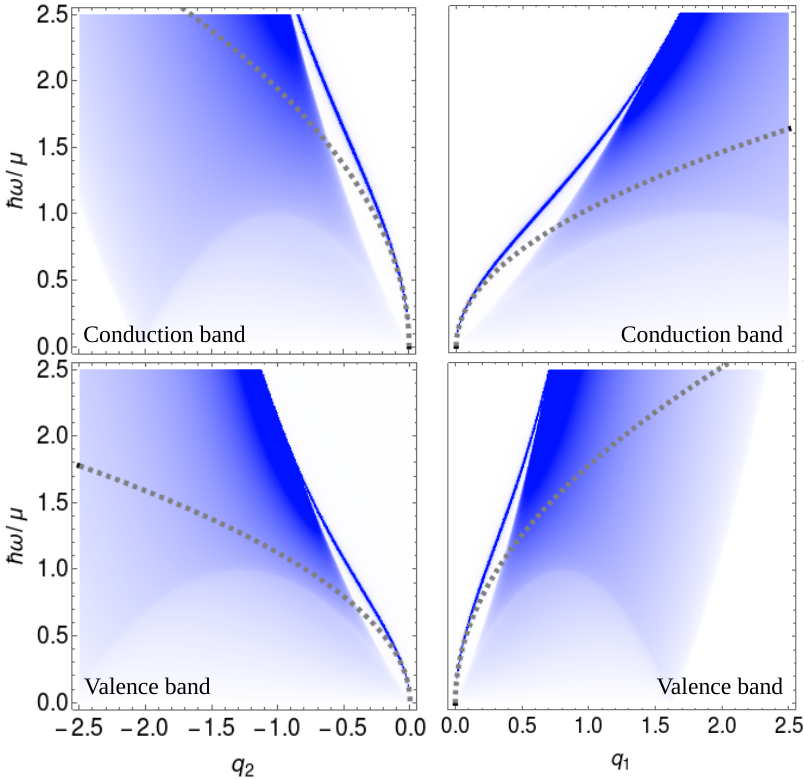}
\caption{Loss function for single layer TiS$_3$ for the valence and conduction bands, along the $q_1$ and $q_2$ axes, calculated from the polarization function (\ref{Eq:Pi0}) at $T=0$. The dashed lines correspond to the analytic approximation (\ref{Eq:Plasmon}).}
\label{Fig:Plasmons-w-q}
\end{figure}

One of the most promising applications of the new families of 2D materials is their potential for optoelectronics and nanoplasmonics.\cite{GPN12,Low-NM16} 
Plasmons are collective density oscillations of an electron liquid that occur in many metals and semiconductors. 
Because of their single or few layer structure, these collective modes in 2D materials are highly confined, and can be tuned and manipulated by external gate or chemical doping. Plasmons have been studied in different 2D materials like graphene,\cite{Shung-PRB86,Wunsch-NJP06,Hwang-PRB07,Stauber-JPCM14,Roldan-PRB13} silicene,\cite{Tabert-PRB14} TMDs,\cite{Scholz-PRB13,Groenewald-PRB16} or black phosphorus.\cite{Low-PRL14} 
In this section we study the plasmon spectrum of doped single-layer TiS$_3$.
The zeros of the dynamical dielectric function $\varepsilon({\bm q},\omega)$ yield the excitation spectrum of the plasmon modes of the electron liquid. The loss function, defined as ${\cal L}(\bq,\omega)=-\Im[1/ \varepsilon(\bq,\omega)]$, quantifies the spectral weight of the plasmon mode. In the limit of zero damping, ${\cal L}(\bq,\omega)$ consists in a delta peak.
The dielectric function can be calculated, within the random phase approximation (RPA) as\cite{AFS82,GV05}
\begin{equation}
\varepsilon(\bq,\omega)=1-V({\bq})\Pi^0({\bq},\omega)
\end{equation}
where $V({\bm q})$ is the Coulomb interaction
\begin{equation}
V({\bq})=\frac{2\pi e^2}{\epsilon_Bq}
\end{equation}
where $\epsilon_B$ is the background dielectric constant and $\Pi^0({\bf q},\omega)$ is the polarization function
\begin{equation}\label{Eq:Pi0}
\Pi^0({\bq},\omega)=\frac{g_s}{{\cal V}}\sum_{{\bk}}\frac{f_F({\bk})-f_F({\bk}+{\bq})}{\xi({\bk}+{\bq})-\xi({\bk})-\hbar(\omega+i\eta)}
\end{equation}
where $g_s=2$ is the spin degeneracy, $\cal V$ is the system size, $f_F({\bk})=\{\exp[\xi(\bk)/k_BT]+1\}^{-1}$ is the Fermi-Dirac distribution function, and $\eta$ is a phenomenological broadening. Here we are interested on low energy intra-band plasmon modes. Therefore, we neglect inter-band transitions in the polarization  (\ref{Eq:Pi0}) and consider the energy dispersion within the effective mass approximation:
\begin{equation}
\xi^{c,v}({\bk})=\frac{\hbar^2k_1^2}{2m^{c,v}_1}+\frac{\hbar^2k_2^2}{2m^{c,v}_2} -\mu
\end{equation}
where $\mu$ is the chemical potential, the superscript $c(v)$ indicates conduction (valence) band, and the corresponding effective masses have been obtained from the DFT band structure and given in Table \ref{tab:my_label}.

As we have seen in the previous sections, single layer TiS$_3$ is a highly anisotropic material, with energy bands dispersing very differently in the two crystallographic directions. Therefore, we follow the scheme used by Low {\it et al.} to study plasmons in black phosphorus,\cite{Low-PRL14} which is another anisotropic 2D crystal that consists in P atoms arranged in a puckered honeycomb lattice. This procedure has been later generalized to study plasmons in rotated double layer systems,\cite{Rodin-PRB15} Coulomb drag in anisotropic van der Waals heterostructures\cite{Pouya-JPCM16} and 2D systems with merging Dirac points.\cite{Pyatkovskiy-PRB16} Denoting ${\bm q}=q(\cos\theta,\sin\theta)$, we can write the polarization function as
\begin{widetext}
\begin{equation}
\Pi^0({\bq},\omega)=\frac{g_{\rm 2D}}{2}\sum_{j=\pm 1}\left(\prod_{l=\pm 1} \sqrt{1+j\frac{\hbar(\omega+i\eta)}{\frac{\hbar^2q^2f(\theta)}{2m_1}} -2l\sqrt{\frac{\mu}{\frac{\hbar^2q^2f(\theta)}{2m_1}}}} -1 \right)
\end{equation}
\end{widetext}
where $g_{\rm 2D}=m_d/\pi\hbar^2$ is the 2D density of states (DOS), in terms of the DOS mass $m_d=\sqrt{m_1m_2}$, and 
\begin{equation}
f(\theta)=\cos^2(\theta)+\frac{m_1}{m_2}\sin^2(\theta).
\end{equation}

\begin{figure}
\centering
\includegraphics[width=80mm]{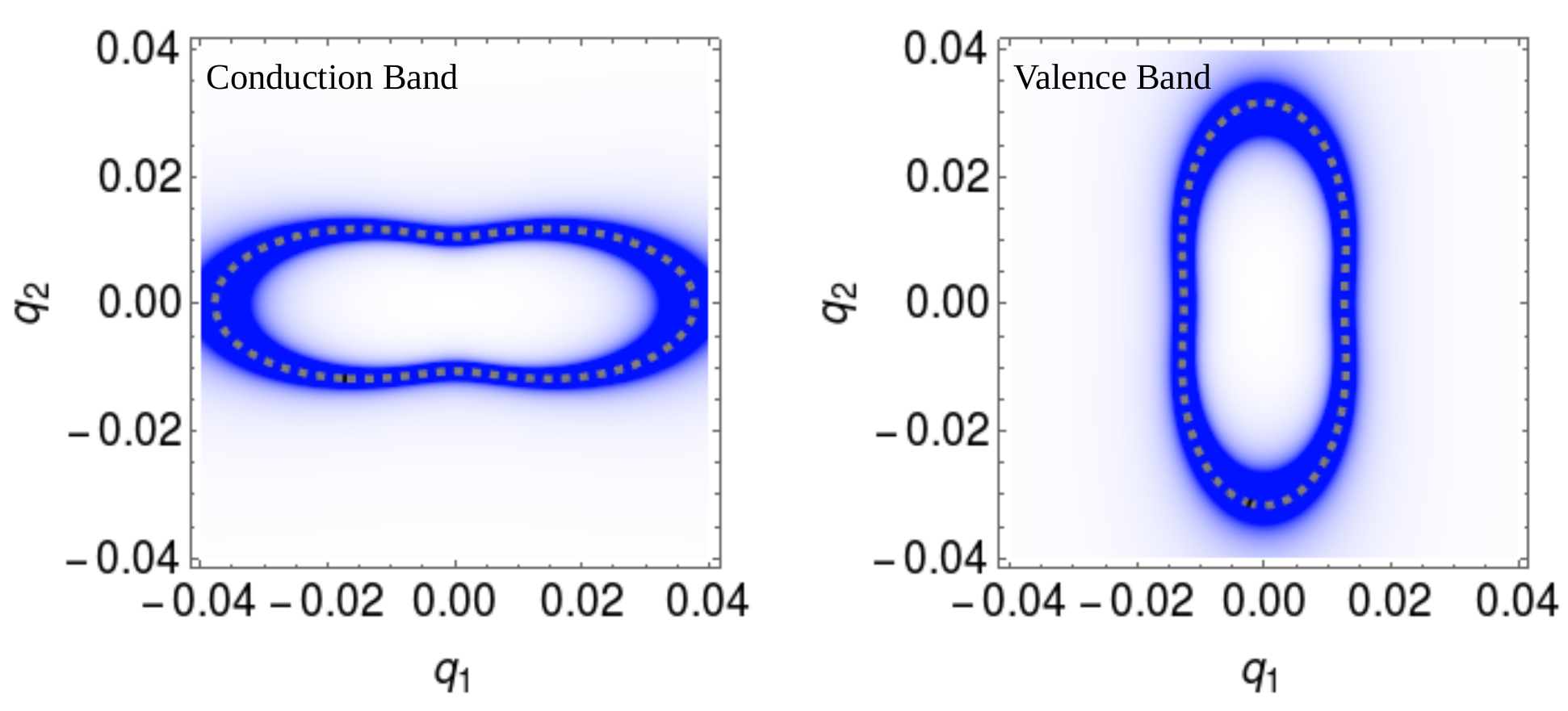}
\caption{Cross section of the loss function ${\cal L}(\bq,\omega)$ that show the plasmon dispersion in single layer TiS$_3$ for the valence and conduction bands in the $q_1-q_2$ plane, for $\hbar\omega/\mu=0.2$ and $T=0$. The dashed lines correspond to the analytic approximation (\ref{Eq:Plasmon}).}
\label{Fig:Plasmons-Polar}
\end{figure}

Our results are shown in Fig. \ref{Fig:Plasmons-w-q}, where we plot the loss function ${\cal L}(\bq,\omega)$ for the valence and conduction bands along the two crystallographic directions. As in the case of black phosphorus,\cite{Low-PRL14,Jin-PRB15} we observe a clear anisotropy in the plasmon mode. However, contrary to black phosphorus, the plasmon anisotropy in TiS$_3$ is opposite for the valence and conduction bands. Indeed, one can clearly observe that for electron (hole) doping, the plasmon dispersion is faster (slower) for wave-vectors $\bq \parallel {\hat{\bm q}}_2$ ($\bq \parallel {\hat{\bm q}}_1$), as seen by the slope of the corresponding modes in Fig. \ref{Fig:Plasmons-w-q}. The anisotropy is more clearly seen in Fig. \ref{Fig:Plasmons-Polar} where we show a cross section of the loss function in the $q_1-q_2$ plane for $\hbar\omega/\mu=0.2$. The origin of the opposite anisotropy for the valence and conduction bands originate from the opposite rate between the effective masses $m_1/m_2$ for each band (see Table \ref{tab:my_label}). This can be well understood from the different orbital character of the valence and conduction band edges, as explained in Sec. \ref{Sec:Nature}. In brief, the top of the valence band is mostly made of the 3$p_x$ orbitals of S coming only from the inner sulfur atoms. This leads to a valence band dispersing mostly along the $q_1-$direction ($m^v_1<m^v_2$). On the other hand, the conduction band is mainly based on the $d_{z^2}$ orbitals of Ti directed along the $q_2-$direction. Therefore the effective masses of the electron carriers are anisotropic with $m^c_1>m^c_2$. This opposite rate between the effective masses for the valence and conduction bands leads at the end of the day to the different behaviour of the collective plasmon modes that appear after electron or hole doping.

\begin{figure}
\centering
\includegraphics[width=80mm]{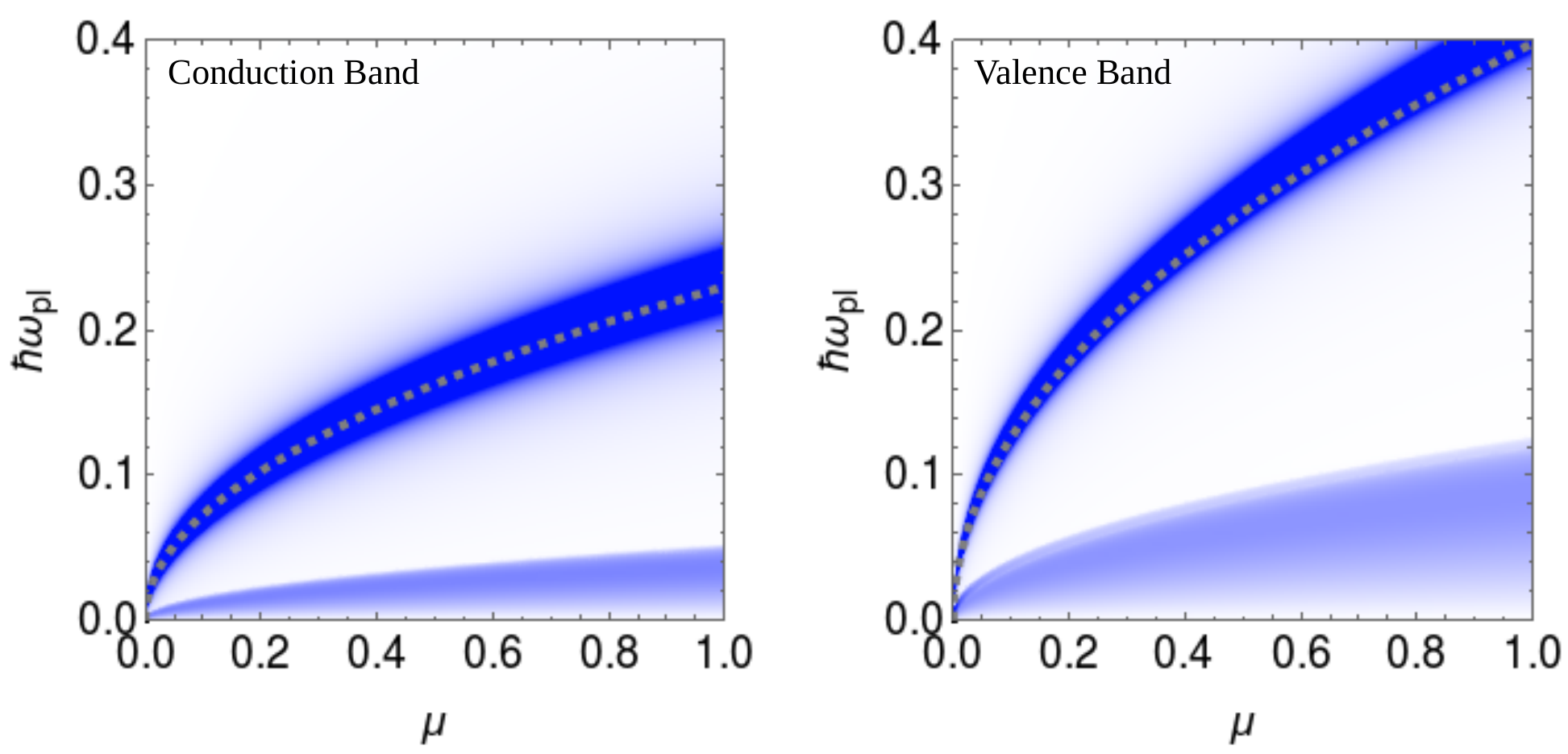}
\caption{Plasmon energies $\hbar \omega_{pl}$ as a function of chemical potential $\mu$, calculated for a single layer TiS$_3$ at a specific $q=0.05a^{-1}$ along the $a$ crystallographic direction, for the conduction and valence bands, obtained from the loss function ${\cal L}(\bq,\omega)$. The dashed lines correspond to the analytic approximation (\ref{Eq:Plasmon}).}
\label{Fig:Plasmons-mu}
\end{figure}

The plasmon dispersion at long wavelengths presents the standard $\sim\sqrt{q}$ behaviour typical for 2D electron gases,\cite{AFS82} with a correction that accounts for the anisotropy\cite{Low-PRL14,Rodin-PRB15,Pouya-JPCM16,Pyatkovskiy-PRB16} and it can be well approximated by the analytical expression
\begin{widetext}
\begin{equation}\label{Eq:Plasmon}
\hbar\omega_{\rm pl}(\bq)=\sqrt{\frac{g_se^2\mu }{\varepsilon_b}\left[\left(\frac{m_2}{m_1}\right)^{1/2}\cos^2\theta+\left(\frac{m_1}{m_2}\right)^{1/2}\sin^2\theta\right]q}.
\end{equation}
\end{widetext}
The dispersion obtained from the approximation (\ref{Eq:Plasmon}) is shown by dashed grey lines in Figs. \ref{Fig:Plasmons-w-q} and \ref{Fig:Plasmons-Polar}. One should notice that the the plasmons are coherent outside the particle-hole continuum, which is defined as the region where $\Im [\Pi^0(\bq,\omega)]\ne 0$. The boundaries of the continuum are defined by 
\begin{equation}
\hbar \omega_{\pm}=\frac{\hbar^2q^2}{2m_1}f(\theta)\pm 2\sqrt{\frac{\hbar^2q^2}{2m_1}\mu f(\theta)}.
\end{equation} 
When the plasmon touches the continuum threshold, the mode starts to be damped by decaying into electron-hole pairs. One of the consequences of the anisotropy of the spectrum is that the damping start to occur at different $\omega-\bq$ rates depending on the direction of propagation.  We finally notice that the plasmon frequency presents the standard $\sqrt{\mu}$ dependence with the chemical potential (see Fig. \ref{Fig:Plasmons-mu}), as expected for a plasmon in a 2D crystal, like in the well known case of graphene.\cite{Shung-PRB86,Wunsch-NJP06,Hwang-PRB07} However, contrary to graphene for which $\mu\propto n^{1/2}$ where $n$ is the carrier density, in TiS$_3$ we have $\mu\propto n$, leading to the expected scaling relation $\omega_{pl}\propto n^{1/2}$, as in standard 2D electron gases,\cite{GV05} while $\omega_{pl}\propto n^{1/4}$ in graphene due its low energy linear dispersion relation.

\section{Conclusions}
In summary, we have used first-principles methods to study the electronic properties of single layer TiS$_3$. Careful analysis of the DFT band structure points out the possibility of changing their physical behaviour by injecting electron carriers. A study of the Fermi surface and Lindhard response function for different electron doping levels shows that for electron doping levels $x$ $\sim0.18-0.30e^-$ ($x$ being the number of electrons per unit cell) the system could exhibit incommensurate charge or spin modulations which, however, would keep the metallic state. In contrast, systems doped with smaller $x$ would be 2D metals without any electronic instability. The inter-chain coupling between the different trigonal prismatic TiS$_3$ chains influences the bottom states of the conduction bands throughout a complex mixing of Ti 3$d_{z^2}$, 3$d_{xy}$ and 3$d_{x^2-y^2}$ orbitals as well as 3$p_x$ orbitals of the bonded S-S pairs. This leads to an unexpected shape of the Fermi surface and Lindhard response function for a large part of the carrier concentrations examined. We further find that, while spin-orbit interaction does not produce any significant modification of the valence and conduction band states, it leads to several avoided band crossings in the spectrum. Finally, we discuss the main features of the plasmon spectrum in doped single layer TiS$_3$ samples. By using a low energy effective mass theory model, we find that plasmons in TiS$_3$ are highly anisotropic. Interestingly, the anisotropy is opposite for the electron and hole branches, which can be understood from the different orbital nature of the valence and conduction bands. We find that the mass anisotropy ratio is $\sim 0.4$ for the valence band and $\sim 3.5$ for the conduction band. This anisotropy, which is opposite for the valence and conduction bands, might be compared to the same directional anisotropy of black phosphorus, with ratios of $\sim 6.2$ and $\sim 5.1$ for the valence and conduction bands respectively.\cite{Yuan_PRB_2015} Another measure of the anisotropy can be obtained from the mobility ratios. In this case, it has  been experimentally shown that few layers TiS$_3$ present a highly temperature dependent ratio of  $\sim 2.3$ to $7.6$ from room temperature to 25~K respectively,\cite{Island_SR_2016} while the anisotropy in the mobility of black phosphorus is of $\sim 1.5$ and only weakly dependent on temperature.\cite{Xia_NC_2014}

Our results show that TiS$_3$ nanofilms are promising platforms for future optoelectronics and nanoelectronics applications, including field effect transistors or infrared photodetectors. In particular, the strong anisotropy of the electrical and optical properties of this material can lead to novel functionalities for devices based on TiS$_3$ like high-performance transistors built along the light effective mass direction, or the directional focus of photocarriers, as demonstrated in black phosphorus,\cite{Quereda_NL_2016} but with the additional feasibility in TiS$_3$ to reverse the main direction of propagation by switching from $n-$ to $p-$doping and viceversa.

\section{Acknowledgments}
We thank A. Castellanos-Gomez for useful discussions and R. Robles for assistance with the calculations. This work has received funding from the European Union’s Seventh Framework Programme (FP7/2007-2013) through the ERC Advanced Grant NOVGRAPHENE (GA No. 290846), European Commission under the Graphene Flagship, contract CNECTICT-604391, the Spanish MINECO (through Grants No. FIS2015-64886-C5-3-P, FIS2015-64886-C5-4-P, FIS2014-58445-JIN and the Severo Ochoa Centers of Excellence Program under Grants SEV-2013-0295 and SEV-2015-0496), and the Generalitat de Catalunya (2014SGR301), as well as from the Comunidad Aut\'onoma de Madrid (CAM) MAD2D-CM Program (S2013/MIT-3007).

\bibliography{TiS3,DFT}

\begin{thebibliography}{64}
\expandafter\ifx\csname natexlab\endcsname\relax\def\natexlab#1{#1}\fi
\expandafter\ifx\csname bibnamefont\endcsname\relax
  \def\bibnamefont#1{#1}\fi
\expandafter\ifx\csname bibfnamefont\endcsname\relax
  \def\bibfnamefont#1{#1}\fi
\expandafter\ifx\csname citenamefont\endcsname\relax
  \def\citenamefont#1{#1}\fi
\expandafter\ifx\csname url\endcsname\relax
  \def\url#1{\texttt{#1}}\fi
\expandafter\ifx\csname urlprefix\endcsname\relax\def\urlprefix{URL }\fi
\providecommand{\bibinfo}[2]{#2}
\providecommand{\eprint}[2][]{\url{#2}}

\bibitem[{\citenamefont{Castro~Neto et~al.}(2009)\citenamefont{Castro~Neto,
  Guinea, Peres, Novoselov, and Geim}}]{review_graphene_2009}
\bibinfo{author}{\bibfnamefont{A.~H.} \bibnamefont{Castro~Neto}},
  \bibinfo{author}{\bibfnamefont{F.}~\bibnamefont{Guinea}},
  \bibinfo{author}{\bibfnamefont{N.~M.~R.} \bibnamefont{Peres}},
  \bibinfo{author}{\bibfnamefont{K.~S.} \bibnamefont{Novoselov}},
  \bibnamefont{and} \bibinfo{author}{\bibfnamefont{A.~K.} \bibnamefont{Geim}},
  \bibinfo{journal}{Reviews of Modern Physics} \textbf{\bibinfo{volume}{81}},
  \bibinfo{pages}{109} (\bibinfo{year}{2009}).

\bibitem[{\citenamefont{Cao et~al.}(2015)\citenamefont{Cao, Mishchenko, Yu,
  Khestanova, Rooney, Prestat, Kretinin, Blake, Shalom, Woods
  et~al.}}]{geim_SC_2015}
\bibinfo{author}{\bibfnamefont{Y.}~\bibnamefont{Cao}},
  \bibinfo{author}{\bibfnamefont{A.}~\bibnamefont{Mishchenko}},
  \bibinfo{author}{\bibfnamefont{G.~L.} \bibnamefont{Yu}},
  \bibinfo{author}{\bibfnamefont{E.}~\bibnamefont{Khestanova}},
  \bibinfo{author}{\bibfnamefont{A.~P.} \bibnamefont{Rooney}},
  \bibinfo{author}{\bibfnamefont{E.}~\bibnamefont{Prestat}},
  \bibinfo{author}{\bibfnamefont{A.~V.} \bibnamefont{Kretinin}},
  \bibinfo{author}{\bibfnamefont{P.}~\bibnamefont{Blake}},
  \bibinfo{author}{\bibfnamefont{M.~B.} \bibnamefont{Shalom}},
  \bibinfo{author}{\bibfnamefont{C.}~\bibnamefont{Woods}},
  \bibnamefont{et~al.}, \bibinfo{journal}{Nano Letters}
  \textbf{\bibinfo{volume}{15}}, \bibinfo{pages}{4914} (\bibinfo{year}{2015}).

\bibitem[{\citenamefont{Mak et~al.}(2010)\citenamefont{Mak, Lee, Hone, Shan,
  and Heinz}}]{mak_mos2_2010}
\bibinfo{author}{\bibfnamefont{K.~F.} \bibnamefont{Mak}},
  \bibinfo{author}{\bibfnamefont{C.}~\bibnamefont{Lee}},
  \bibinfo{author}{\bibfnamefont{J.}~\bibnamefont{Hone}},
  \bibinfo{author}{\bibfnamefont{J.}~\bibnamefont{Shan}}, \bibnamefont{and}
  \bibinfo{author}{\bibfnamefont{T.~F.} \bibnamefont{Heinz}},
  \bibinfo{journal}{Physical Review Letters} \textbf{\bibinfo{volume}{105}},
  \bibinfo{pages}{136805} (\bibinfo{year}{2010}).

\bibitem[{\citenamefont{Novoselov et~al.}(2005)\citenamefont{Novoselov, Jiang,
  Schedin, Booth, Khotkevich, Morozov, and Geim}}]{Geim_PNAS_2005}
\bibinfo{author}{\bibfnamefont{K.~S.} \bibnamefont{Novoselov}},
  \bibinfo{author}{\bibfnamefont{D.}~\bibnamefont{Jiang}},
  \bibinfo{author}{\bibfnamefont{F.}~\bibnamefont{Schedin}},
  \bibinfo{author}{\bibfnamefont{T.~J.} \bibnamefont{Booth}},
  \bibinfo{author}{\bibfnamefont{V.~V.} \bibnamefont{Khotkevich}},
  \bibinfo{author}{\bibfnamefont{S.~V.} \bibnamefont{Morozov}},
  \bibnamefont{and} \bibinfo{author}{\bibfnamefont{A.~K.} \bibnamefont{Geim}},
  \bibinfo{journal}{Proc. Natl. Acad. Sci USA} \textbf{\bibinfo{volume}{102}},
  \bibinfo{pages}{10451} (\bibinfo{year}{2005}).

\bibitem[{\citenamefont{Wang et~al.}(2012)\citenamefont{Wang, Kalantar-Zadeh,
  Kis, Coleman, and Strano}}]{Review_Wang_2012}
\bibinfo{author}{\bibfnamefont{Q.~H.} \bibnamefont{Wang}},
  \bibinfo{author}{\bibfnamefont{K.}~\bibnamefont{Kalantar-Zadeh}},
  \bibinfo{author}{\bibfnamefont{A.}~\bibnamefont{Kis}},
  \bibinfo{author}{\bibfnamefont{J.~N.} \bibnamefont{Coleman}},
  \bibnamefont{and} \bibinfo{author}{\bibfnamefont{M.~S.}
  \bibnamefont{Strano}}, \bibinfo{journal}{Nature Nanotechnology}
  \textbf{\bibinfo{volume}{7}}, \bibinfo{pages}{699} (\bibinfo{year}{2012}).

\bibitem[{\citenamefont{Rold\'an et~al.}(2014)\citenamefont{Rold\'an,
  Silva-Guill\'en, L\'opez-Sancho, Guinea, Cappelluti, and
  Ordej\'on}}]{roldan_review_2014}
\bibinfo{author}{\bibfnamefont{R.}~\bibnamefont{Rold\'an}},
  \bibinfo{author}{\bibfnamefont{J.~A.} \bibnamefont{Silva-Guill\'en}},
  \bibinfo{author}{\bibfnamefont{M.~P.} \bibnamefont{L\'opez-Sancho}},
  \bibinfo{author}{\bibfnamefont{F.}~\bibnamefont{Guinea}},
  \bibinfo{author}{\bibfnamefont{E.}~\bibnamefont{Cappelluti}},
  \bibnamefont{and}
  \bibinfo{author}{\bibfnamefont{P.}~\bibnamefont{Ordej\'on}},
  \bibinfo{journal}{Annalen der Physik} \textbf{\bibinfo{volume}{526}},
  \bibinfo{pages}{347} (\bibinfo{year}{2014}).

\bibitem[{\citenamefont{Guo et~al.}(2014)\citenamefont{Guo, Lu, Wang, Wu, and
  Zeng}}]{Guo_2014}
\bibinfo{author}{\bibfnamefont{H.}~\bibnamefont{Guo}},
  \bibinfo{author}{\bibfnamefont{N.}~\bibnamefont{Lu}},
  \bibinfo{author}{\bibfnamefont{L.}~\bibnamefont{Wang}},
  \bibinfo{author}{\bibfnamefont{X.}~\bibnamefont{Wu}}, \bibnamefont{and}
  \bibinfo{author}{\bibfnamefont{X.~C.} \bibnamefont{Zeng}},
  \bibinfo{journal}{J. Phys. Chem. C} \textbf{\bibinfo{volume}{118}},
  \bibinfo{pages}{7242} (\bibinfo{year}{2014}).

\bibitem[{\citenamefont{Zhou et~al.}(2012)\citenamefont{Zhou, Wang, Yang, Zu,
  Yang, Sun, and Gao}}]{Zhou_2012}
\bibinfo{author}{\bibfnamefont{Y.}~\bibnamefont{Zhou}},
  \bibinfo{author}{\bibfnamefont{Z.}~\bibnamefont{Wang}},
  \bibinfo{author}{\bibfnamefont{P.}~\bibnamefont{Yang}},
  \bibinfo{author}{\bibfnamefont{X.}~\bibnamefont{Zu}},
  \bibinfo{author}{\bibfnamefont{L.}~\bibnamefont{Yang}},
  \bibinfo{author}{\bibfnamefont{X.}~\bibnamefont{Sun}}, \bibnamefont{and}
  \bibinfo{author}{\bibfnamefont{F.}~\bibnamefont{Gao}}, \bibinfo{journal}{ACS
  Nano} \textbf{\bibinfo{volume}{6}}, \bibinfo{pages}{9727}
  (\bibinfo{year}{2012}).

\bibitem[{\citenamefont{R\"osner et~al.}(2014)\citenamefont{R\"osner, Haas, and
  Wehling}}]{Rosner_2014}
\bibinfo{author}{\bibfnamefont{M.}~\bibnamefont{R\"osner}},
  \bibinfo{author}{\bibfnamefont{S.}~\bibnamefont{Haas}}, \bibnamefont{and}
  \bibinfo{author}{\bibfnamefont{T.~O.} \bibnamefont{Wehling}},
  \bibinfo{journal}{Phys. Rev. B} \textbf{\bibinfo{volume}{90}},
  \bibinfo{pages}{245105} (\bibinfo{year}{2014}).

\bibitem[{\citenamefont{Dai et~al.}(2016)\citenamefont{Dai, Li, and
  Zeng}}]{dai_2016}
\bibinfo{author}{\bibfnamefont{J.}~\bibnamefont{Dai}},
  \bibinfo{author}{\bibfnamefont{M.}~\bibnamefont{Li}}, \bibnamefont{and}
  \bibinfo{author}{\bibfnamefont{X.~C.} \bibnamefont{Zeng}},
  \bibinfo{journal}{WIREs Computational Molecular Science}
  \textbf{\bibinfo{volume}{6}}, \bibinfo{pages}{211} (\bibinfo{year}{2016}).

\bibitem[{\citenamefont{Island et~al.}(2017)\citenamefont{Island,
  Molina-Mendoza, Barawi, Biele, Flores, Clamagirand, Ares, Sanchez, van~der
  Zant, D'Agosta et~al.}}]{Island2017Electronics}
\bibinfo{author}{\bibfnamefont{J.~O.} \bibnamefont{Island}},
  \bibinfo{author}{\bibfnamefont{A.~J.} \bibnamefont{Molina-Mendoza}},
  \bibinfo{author}{\bibfnamefont{M.}~\bibnamefont{Barawi}},
  \bibinfo{author}{\bibfnamefont{R.}~\bibnamefont{Biele}},
  \bibinfo{author}{\bibfnamefont{E.}~\bibnamefont{Flores}},
  \bibinfo{author}{\bibfnamefont{J.~M.} \bibnamefont{Clamagirand}},
  \bibinfo{author}{\bibfnamefont{J.~R.} \bibnamefont{Ares}},
  \bibinfo{author}{\bibfnamefont{C.}~\bibnamefont{Sanchez}},
  \bibinfo{author}{\bibfnamefont{H.~S.~J.} \bibnamefont{van~der Zant}},
  \bibinfo{author}{\bibfnamefont{R.}~\bibnamefont{D'Agosta}},
  \bibnamefont{et~al.}, \bibinfo{journal}{arXiv}  (\bibinfo{year}{2017}),
  \eprint{1702.01865}.

\bibitem[{\citenamefont{Island et~al.}(2015)\citenamefont{Island, Barawi,
  Biele, Almazán, Clamagirand, Ares, S\'anchez, van~der Zant, \'Alvarez,
  D'Agosta et~al.}}]{TiS3_Castellanos_2015}
\bibinfo{author}{\bibfnamefont{J.~O.} \bibnamefont{Island}},
  \bibinfo{author}{\bibfnamefont{M.}~\bibnamefont{Barawi}},
  \bibinfo{author}{\bibfnamefont{R.}~\bibnamefont{Biele}},
  \bibinfo{author}{\bibfnamefont{A.}~\bibnamefont{Almazán}},
  \bibinfo{author}{\bibfnamefont{J.~M.} \bibnamefont{Clamagirand}},
  \bibinfo{author}{\bibfnamefont{J.~R.} \bibnamefont{Ares}},
  \bibinfo{author}{\bibfnamefont{C.}~\bibnamefont{S\'anchez}},
  \bibinfo{author}{\bibfnamefont{H.~S.~J.} \bibnamefont{van~der Zant}},
  \bibinfo{author}{\bibfnamefont{J.~V.} \bibnamefont{\'Alvarez}},
  \bibinfo{author}{\bibfnamefont{R.}~\bibnamefont{D'Agosta}},
  \bibnamefont{et~al.}, \bibinfo{journal}{Advanced Materials}
  \textbf{\bibinfo{volume}{27}}, \bibinfo{pages}{2595} (\bibinfo{year}{2015}).

\bibitem[{\citenamefont{Jin et~al.}(2015{\natexlab{a}})\citenamefont{Jin, Li,
  and Yang}}]{jin_2015}
\bibinfo{author}{\bibfnamefont{Y.}~\bibnamefont{Jin}},
  \bibinfo{author}{\bibfnamefont{X.}~\bibnamefont{Li}}, \bibnamefont{and}
  \bibinfo{author}{\bibfnamefont{J.}~\bibnamefont{Yang}},
  \bibinfo{journal}{Phys. Chem. Chem. Phys.} \textbf{\bibinfo{volume}{17}},
  \bibinfo{pages}{18665} (\bibinfo{year}{2015}{\natexlab{a}}).

\bibitem[{\citenamefont{Bullett}(1979)}]{Bullett_1979}
\bibinfo{author}{\bibfnamefont{D.~W.} \bibnamefont{Bullett}},
  \bibinfo{journal}{Journal of Physics C: Solid State Physics}
  \textbf{\bibinfo{volume}{12}}, \bibinfo{pages}{277} (\bibinfo{year}{1979}).

\bibitem[{\citenamefont{Myron et~al.}(1981)\citenamefont{Myron, Harmon, and
  Khumalo}}]{Myron_1981}
\bibinfo{author}{\bibfnamefont{H.~W.} \bibnamefont{Myron}},
  \bibinfo{author}{\bibfnamefont{B.~N.} \bibnamefont{Harmon}},
  \bibnamefont{and} \bibinfo{author}{\bibfnamefont{F.}~\bibnamefont{Khumalo}},
  \bibinfo{journal}{Journal of Physics and Chemistry of Solids}
  \textbf{\bibinfo{volume}{42}}, \bibinfo{pages}{263} (\bibinfo{year}{1981}).

\bibitem[{\citenamefont{Canadell et~al.}(1988)\citenamefont{Canadell, Mathey,
  and Whangbo}}]{Canadell_1988}
\bibinfo{author}{\bibfnamefont{E.}~\bibnamefont{Canadell}},
  \bibinfo{author}{\bibfnamefont{Y.}~\bibnamefont{Mathey}}, \bibnamefont{and}
  \bibinfo{author}{\bibfnamefont{M.-H.} \bibnamefont{Whangbo}},
  \bibinfo{journal}{Journal of the American Chemical Society}
  \textbf{\bibinfo{volume}{110}}, \bibinfo{pages}{104} (\bibinfo{year}{1988}).

\bibitem[{\citenamefont{Canadell et~al.}(1989)\citenamefont{Canadell, Thieffry,
  Mathey, and Whangbo}}]{Canadell_1989}
\bibinfo{author}{\bibfnamefont{E.}~\bibnamefont{Canadell}},
  \bibinfo{author}{\bibfnamefont{C.}~\bibnamefont{Thieffry}},
  \bibinfo{author}{\bibfnamefont{Y.}~\bibnamefont{Mathey}}, \bibnamefont{and}
  \bibinfo{author}{\bibfnamefont{M.-H.} \bibnamefont{Whangbo}},
  \bibinfo{journal}{Inorganic Chemistry} \textbf{\bibinfo{volume}{28}},
  \bibinfo{pages}{3043} (\bibinfo{year}{1989}).

\bibitem[{\citenamefont{Stowe and Wagner}(1998)}]{Stowe_1998}
\bibinfo{author}{\bibfnamefont{K.}~\bibnamefont{Stowe}} \bibnamefont{and}
  \bibinfo{author}{\bibfnamefont{F.~R.} \bibnamefont{Wagner}},
  \bibinfo{journal}{Journal of Solid State Chemistry}
  \textbf{\bibinfo{volume}{138}}, \bibinfo{pages}{160} (\bibinfo{year}{1998}).

\bibitem[{\citenamefont{Felser et~al.}(1998)\citenamefont{Felser, Finckh,
  Kleinke, Rocker, and Tremel}}]{Felser_1998}
\bibinfo{author}{\bibfnamefont{C.}~\bibnamefont{Felser}},
  \bibinfo{author}{\bibfnamefont{E.}~\bibnamefont{Finckh}},
  \bibinfo{author}{\bibfnamefont{H.}~\bibnamefont{Kleinke}},
  \bibinfo{author}{\bibfnamefont{F.}~\bibnamefont{Rocker}}, \bibnamefont{and}
  \bibinfo{author}{\bibfnamefont{W.}~\bibnamefont{Tremel}},
  \bibinfo{journal}{Journal of Materials Chemistry}
  \textbf{\bibinfo{volume}{8}}, \bibinfo{pages}{1787} (\bibinfo{year}{1998}).

\bibitem[{\citenamefont{Dai and Zeng}(2015)}]{dai_2015}
\bibinfo{author}{\bibfnamefont{J.}~\bibnamefont{Dai}} \bibnamefont{and}
  \bibinfo{author}{\bibfnamefont{X.~C.} \bibnamefont{Zeng}},
  \bibinfo{journal}{Angewandte Chemie International Edition}
  \textbf{\bibinfo{volume}{54}}, \bibinfo{pages}{7572} (\bibinfo{year}{2015}).

\bibitem[{\citenamefont{Li et~al.}(2015)\citenamefont{Li, Dai, and
  Zeng}}]{Li_Nano_2015}
\bibinfo{author}{\bibfnamefont{M.}~\bibnamefont{Li}},
  \bibinfo{author}{\bibfnamefont{J.}~\bibnamefont{Dai}}, \bibnamefont{and}
  \bibinfo{author}{\bibfnamefont{X.~C.} \bibnamefont{Zeng}},
  \bibinfo{journal}{Nanoscale} \textbf{\bibinfo{volume}{7}},
  \bibinfo{pages}{15385} (\bibinfo{year}{2015}).

\bibitem[{\citenamefont{Iyikanat et~al.}(2015)\citenamefont{Iyikanat, Sahin,
  Senger, and Peeters}}]{Peeters_2015}
\bibinfo{author}{\bibfnamefont{F.}~\bibnamefont{Iyikanat}},
  \bibinfo{author}{\bibfnamefont{H.}~\bibnamefont{Sahin}},
  \bibinfo{author}{\bibfnamefont{R.~T.} \bibnamefont{Senger}},
  \bibnamefont{and} \bibinfo{author}{\bibfnamefont{F.~M.}
  \bibnamefont{Peeters}}, \bibinfo{journal}{Journal of Physical Chemistry C}
  \textbf{\bibinfo{volume}{119}}, \bibinfo{pages}{10709}
  (\bibinfo{year}{2015}).

\bibitem[{\citenamefont{Kang and Wang}(2016)}]{Kang_Jun_2016}
\bibinfo{author}{\bibfnamefont{K.}~\bibnamefont{Kang}} \bibnamefont{and}
  \bibinfo{author}{\bibfnamefont{L.-W.} \bibnamefont{Wang}},
  \bibinfo{journal}{Phys. Chem. Chem. Phys.} \textbf{\bibinfo{volume}{18}},
  \bibinfo{pages}{14805} (\bibinfo{year}{2016}).

\bibitem[{\citenamefont{Sambongi}(1986)}]{Sambongi_1986}
\bibinfo{author}{\bibfnamefont{T.}~\bibnamefont{Sambongi}}, in
  \emph{\bibinfo{booktitle}{Crystal Chemistry and Properties of Materials with
  Quasi-One-Dimensional Structures}}, edited by
  \bibinfo{editor}{\bibfnamefont{J.}~\bibnamefont{Rouxel}}
  (\bibinfo{publisher}{Reidel: Dordrecht, The Netherlands},
  \bibinfo{year}{1986}), pp. \bibinfo{pages}{281--313}.

\bibitem[{\citenamefont{Takahashi et~al.}(1984)\citenamefont{Takahashi,
  Sambongi, Brill, and Roark}}]{Sambongi_1984}
\bibinfo{author}{\bibfnamefont{S.}~\bibnamefont{Takahashi}},
  \bibinfo{author}{\bibfnamefont{T.}~\bibnamefont{Sambongi}},
  \bibinfo{author}{\bibfnamefont{J.~W.} \bibnamefont{Brill}}, \bibnamefont{and}
  \bibinfo{author}{\bibfnamefont{W.}~\bibnamefont{Roark}},
  \bibinfo{journal}{Solid State Communications} \textbf{\bibinfo{volume}{49}},
  \bibinfo{pages}{1031} (\bibinfo{year}{1984}).

\bibitem[{\citenamefont{Yomo et~al.}(2005)\citenamefont{Yomo, Kamaya, Abliz,
  Hedo, and Uwatoko}}]{Yomo_2005}
\bibinfo{author}{\bibfnamefont{R.}~\bibnamefont{Yomo}},
  \bibinfo{author}{\bibfnamefont{K.}~\bibnamefont{Kamaya}},
  \bibinfo{author}{\bibfnamefont{M.}~\bibnamefont{Abliz}},
  \bibinfo{author}{\bibfnamefont{M.}~\bibnamefont{Hedo}}, \bibnamefont{and}
  \bibinfo{author}{\bibfnamefont{Y.}~\bibnamefont{Uwatoko}},
  \bibinfo{journal}{Physical Review B} \textbf{\bibinfo{volume}{71}},
  \bibinfo{pages}{132508} (\bibinfo{year}{2005}).

\bibitem[{\citenamefont{Hohenberg and Kohn}(1964)}]{HohKoh1964}
\bibinfo{author}{\bibfnamefont{P.}~\bibnamefont{Hohenberg}} \bibnamefont{and}
  \bibinfo{author}{\bibfnamefont{W.}~\bibnamefont{Kohn}},
  \bibinfo{journal}{Physical Review} \textbf{\bibinfo{volume}{136}},
  \bibinfo{pages}{B864} (\bibinfo{year}{1964}).

\bibitem[{\citenamefont{Kohn and Sham}(1965)}]{KohSha1965}
\bibinfo{author}{\bibfnamefont{W.}~\bibnamefont{Kohn}} \bibnamefont{and}
  \bibinfo{author}{\bibfnamefont{L.~J.} \bibnamefont{Sham}},
  \bibinfo{journal}{Physical Review} \textbf{\bibinfo{volume}{140}},
  \bibinfo{pages}{A1133} (\bibinfo{year}{1965}).

\bibitem[{\citenamefont{Soler et~al.}(2002)\citenamefont{Soler, Artacho, Gale,
  Garc\'ia, Junquera, Ordej\'on, and S\'anchez-Portal}}]{SolArt2002}
\bibinfo{author}{\bibfnamefont{J.~M.} \bibnamefont{Soler}},
  \bibinfo{author}{\bibfnamefont{E.}~\bibnamefont{Artacho}},
  \bibinfo{author}{\bibfnamefont{J.~D.} \bibnamefont{Gale}},
  \bibinfo{author}{\bibfnamefont{A.}~\bibnamefont{Garc\'ia}},
  \bibinfo{author}{\bibfnamefont{J.}~\bibnamefont{Junquera}},
  \bibinfo{author}{\bibfnamefont{P.}~\bibnamefont{Ordej\'on}},
  \bibnamefont{and}
  \bibinfo{author}{\bibfnamefont{D.}~\bibnamefont{S\'anchez-Portal}},
  \bibinfo{journal}{Journal of Physics: Condensed Matter}
  \textbf{\bibinfo{volume}{14}}, \bibinfo{pages}{2745} (\bibinfo{year}{2002}).

\bibitem[{\citenamefont{Artacho et~al.}(2008)\citenamefont{Artacho, Anglada,
  Di\'eguez, Gale, Garc\'ia, Junquera, Martin, Ordej\'on, Pruneda,
  S\'anchez-Portal et~al.}}]{ArtAng2008}
\bibinfo{author}{\bibfnamefont{E.}~\bibnamefont{Artacho}},
  \bibinfo{author}{\bibfnamefont{E.}~\bibnamefont{Anglada}},
  \bibinfo{author}{\bibfnamefont{O.}~\bibnamefont{Di\'eguez}},
  \bibinfo{author}{\bibfnamefont{J.~D.} \bibnamefont{Gale}},
  \bibinfo{author}{\bibfnamefont{A.}~\bibnamefont{Garc\'ia}},
  \bibinfo{author}{\bibfnamefont{J.}~\bibnamefont{Junquera}},
  \bibinfo{author}{\bibfnamefont{R.~M.} \bibnamefont{Martin}},
  \bibinfo{author}{\bibfnamefont{P.}~\bibnamefont{Ordej\'on}},
  \bibinfo{author}{\bibfnamefont{J.~M.} \bibnamefont{Pruneda}},
  \bibinfo{author}{\bibfnamefont{D.}~\bibnamefont{S\'anchez-Portal}},
  \bibnamefont{et~al.}, \bibinfo{journal}{Journal of Physics: Condensed Matter}
  \textbf{\bibinfo{volume}{20}}, \bibinfo{pages}{064208}
  (\bibinfo{year}{2008}).

\bibitem[{\citenamefont{Perdew et~al.}(1996)\citenamefont{Perdew, Burke, and
  Ernzerhof}}]{PBE96}
\bibinfo{author}{\bibfnamefont{J.~P.} \bibnamefont{Perdew}},
  \bibinfo{author}{\bibfnamefont{K.}~\bibnamefont{Burke}}, \bibnamefont{and}
  \bibinfo{author}{\bibfnamefont{M.}~\bibnamefont{Ernzerhof}},
  \bibinfo{journal}{Physical Review Letters} \textbf{\bibinfo{volume}{77}},
  \bibinfo{pages}{3865} (\bibinfo{year}{1996}).

\bibitem[{\citenamefont{Troullier and Martins}(1991)}]{tro91}
\bibinfo{author}{\bibfnamefont{N.}~\bibnamefont{Troullier}} \bibnamefont{and}
  \bibinfo{author}{\bibfnamefont{J.~L.} \bibnamefont{Martins}},
  \bibinfo{journal}{Physical Review B} \textbf{\bibinfo{volume}{43}},
  \bibinfo{pages}{1993} (\bibinfo{year}{1991}).

\bibitem[{\citenamefont{Kleinman and Bylander}(1982)}]{klby82}
\bibinfo{author}{\bibfnamefont{L.}~\bibnamefont{Kleinman}} \bibnamefont{and}
  \bibinfo{author}{\bibfnamefont{D.~M.} \bibnamefont{Bylander}},
  \bibinfo{journal}{Physical Review Letters} \textbf{\bibinfo{volume}{48}},
  \bibinfo{pages}{1425} (\bibinfo{year}{1982}).

\bibitem[{\citenamefont{Louie et~al.}(1982)\citenamefont{Louie, Froyen, and
  Cohen}}]{LFC82}
\bibinfo{author}{\bibfnamefont{S.~G.} \bibnamefont{Louie}},
  \bibinfo{author}{\bibfnamefont{S.}~\bibnamefont{Froyen}}, \bibnamefont{and}
  \bibinfo{author}{\bibfnamefont{M.~L.} \bibnamefont{Cohen}},
  \bibinfo{journal}{Physical Review B} \textbf{\bibinfo{volume}{26}},
  \bibinfo{pages}{1738} (\bibinfo{year}{1982}).

\bibitem[{\citenamefont{Artacho et~al.}(1999)\citenamefont{Artacho,
  S\'anchez-Portal, Ordej\'on, Garc\'ia, and Soler}}]{arsan99}
\bibinfo{author}{\bibfnamefont{E.}~\bibnamefont{Artacho}},
  \bibinfo{author}{\bibfnamefont{D.}~\bibnamefont{S\'anchez-Portal}},
  \bibinfo{author}{\bibfnamefont{P.}~\bibnamefont{Ordej\'on}},
  \bibinfo{author}{\bibfnamefont{A.}~\bibnamefont{Garc\'ia}}, \bibnamefont{and}
  \bibinfo{author}{\bibfnamefont{J.~M.} \bibnamefont{Soler}},
  \bibinfo{journal}{Physica Status Solidi (b)} \textbf{\bibinfo{volume}{215}},
  \bibinfo{pages}{809} (\bibinfo{year}{1999}).

\bibitem[{\citenamefont{Monkhorst and Pack}(1976)}]{MonPac76}
\bibinfo{author}{\bibfnamefont{H.~J.} \bibnamefont{Monkhorst}}
  \bibnamefont{and} \bibinfo{author}{\bibfnamefont{J.~D.} \bibnamefont{Pack}},
  \bibinfo{journal}{Physical Review B} \textbf{\bibinfo{volume}{13}},
  \bibinfo{pages}{5188} (\bibinfo{year}{1976}).

\bibitem[{\citenamefont{Heyd et~al.}(2006)\citenamefont{Heyd, Scuseria, and
  Ernzerhof}}]{HSE06}
\bibinfo{author}{\bibfnamefont{J.}~\bibnamefont{Heyd}},
  \bibinfo{author}{\bibfnamefont{G.~E.} \bibnamefont{Scuseria}},
  \bibnamefont{and}
  \bibinfo{author}{\bibfnamefont{M.}~\bibnamefont{Ernzerhof}},
  \bibinfo{journal}{The Journal of Chemical Physics}
  \textbf{\bibinfo{volume}{124}}, \bibinfo{pages}{219906}
  (\bibinfo{year}{2006}).

\bibitem[{\citenamefont{Kresse and Furthm{\"{u}}ller}(1996)}]{vasp}
\bibinfo{author}{\bibfnamefont{G.}~\bibnamefont{Kresse}} \bibnamefont{and}
  \bibinfo{author}{\bibfnamefont{J.}~\bibnamefont{Furthm{\"{u}}ller}},
  \bibinfo{journal}{Physical Review B} \textbf{\bibinfo{volume}{54}},
  \bibinfo{pages}{11169} (\bibinfo{year}{1996}).

\bibitem[{\citenamefont{Furuseth et~al.}(1975)\citenamefont{Furuseth,
  Bratt{\aa}s, and Kjekshus}}]{furuseth_75}
\bibinfo{author}{\bibfnamefont{S.}~\bibnamefont{Furuseth}},
  \bibinfo{author}{\bibfnamefont{L.}~\bibnamefont{Bratt{\aa}s}},
  \bibnamefont{and} \bibinfo{author}{\bibfnamefont{A.}~\bibnamefont{Kjekshus}},
  \bibinfo{journal}{Acta Chemica Scandinavica A} \textbf{\bibinfo{volume}{29}},
  \bibinfo{pages}{623} (\bibinfo{year}{1975}).

\bibitem[{\citenamefont{Furuseth and Fjellv{\aa}g}(1991)}]{furuseth_91}
\bibinfo{author}{\bibfnamefont{S.}~\bibnamefont{Furuseth}} \bibnamefont{and}
  \bibinfo{author}{\bibfnamefont{H.}~\bibnamefont{Fjellv{\aa}g}},
  \bibinfo{journal}{Acta Chemica Scandinavica} \textbf{\bibinfo{volume}{45}},
  \bibinfo{pages}{694} (\bibinfo{year}{1991}).

\bibitem[{\citenamefont{Grimmeiss et~al.}(1961)\citenamefont{Grimmeiss,
  Rabenau, Hahn, and Ness}}]{opt_gap_1961}
\bibinfo{author}{\bibfnamefont{H.~G.} \bibnamefont{Grimmeiss}},
  \bibinfo{author}{\bibfnamefont{A.}~\bibnamefont{Rabenau}},
  \bibinfo{author}{\bibfnamefont{H.}~\bibnamefont{Hahn}}, \bibnamefont{and}
  \bibinfo{author}{\bibfnamefont{P.}~\bibnamefont{Ness}}, \bibinfo{journal}{Z.
  Elektrochem.} \textbf{\bibinfo{volume}{65}}, \bibinfo{pages}{776}
  (\bibinfo{year}{1961}).

\bibitem[{\citenamefont{Ye et~al.}(2012)\citenamefont{Ye, Zhang, Akashi,
  Bahramy, Arita, and Iwasa}}]{MoS2_doping}
\bibinfo{author}{\bibfnamefont{J.~T.} \bibnamefont{Ye}},
  \bibinfo{author}{\bibfnamefont{Y.~J.} \bibnamefont{Zhang}},
  \bibinfo{author}{\bibfnamefont{R.}~\bibnamefont{Akashi}},
  \bibinfo{author}{\bibfnamefont{M.~S.} \bibnamefont{Bahramy}},
  \bibinfo{author}{\bibfnamefont{R.}~\bibnamefont{Arita}}, \bibnamefont{and}
  \bibinfo{author}{\bibfnamefont{Y.}~\bibnamefont{Iwasa}},
  \bibinfo{journal}{Science} \textbf{\bibinfo{volume}{338}},
  \bibinfo{pages}{1193} (\bibinfo{year}{2012}).

\bibitem[{\citenamefont{Hoffmann et~al.}(1980)\citenamefont{Hoffmann, Shaik,
  Scott, and Foshee}}]{NbSe3}
\bibinfo{author}{\bibfnamefont{R.}~\bibnamefont{Hoffmann}},
  \bibinfo{author}{\bibfnamefont{S.}~\bibnamefont{Shaik}},
  \bibinfo{author}{\bibfnamefont{M.-H.} \bibnamefont{Scott},
  \bibfnamefont{J.~C.~Whangbo}}, \bibnamefont{and}
  \bibinfo{author}{\bibfnamefont{M.~J.} \bibnamefont{Foshee}},
  \bibinfo{journal}{Journal of Solid State Chemistry}
  \textbf{\bibinfo{volume}{34}}, \bibinfo{pages}{263} (\bibinfo{year}{1980}).

\bibitem[{\citenamefont{{Grigorenko} et~al.}(2012)\citenamefont{{Grigorenko},
  {Polini}, and {Novoselov}}}]{GPN12}
\bibinfo{author}{\bibfnamefont{A.~N.} \bibnamefont{{Grigorenko}}},
  \bibinfo{author}{\bibfnamefont{M.}~\bibnamefont{{Polini}}}, \bibnamefont{and}
  \bibinfo{author}{\bibfnamefont{K.~S.} \bibnamefont{{Novoselov}}},
  \bibinfo{journal}{Nat. Photonics} \textbf{\bibinfo{volume}{6}},
  \bibinfo{pages}{749} (\bibinfo{year}{2012}).

\bibitem[{\citenamefont{Low et~al.}(2017)\citenamefont{Low, Chaves, Caldwell,
  Kumar, Fang, Avouris, Heinz, Guinea, Martin-Moreno, and Koppens}}]{Low-NM16}
\bibinfo{author}{\bibfnamefont{T.}~\bibnamefont{Low}},
  \bibinfo{author}{\bibfnamefont{A.}~\bibnamefont{Chaves}},
  \bibinfo{author}{\bibfnamefont{J.~D.} \bibnamefont{Caldwell}},
  \bibinfo{author}{\bibfnamefont{A.}~\bibnamefont{Kumar}},
  \bibinfo{author}{\bibfnamefont{N.~X.} \bibnamefont{Fang}},
  \bibinfo{author}{\bibfnamefont{P.}~\bibnamefont{Avouris}},
  \bibinfo{author}{\bibfnamefont{T.~F.} \bibnamefont{Heinz}},
  \bibinfo{author}{\bibfnamefont{F.}~\bibnamefont{Guinea}},
  \bibinfo{author}{\bibfnamefont{L.}~\bibnamefont{Martin-Moreno}},
  \bibnamefont{and} \bibinfo{author}{\bibfnamefont{F.}~\bibnamefont{Koppens}},
  \bibinfo{journal}{Nature Mat.} \textbf{\bibinfo{volume}{16}},
  \bibinfo{pages}{182} (\bibinfo{year}{2017}).

\bibitem[{\citenamefont{Shung}(1986)}]{Shung-PRB86}
\bibinfo{author}{\bibfnamefont{K.~W.-K.} \bibnamefont{Shung}},
  \bibinfo{journal}{Phys. Rev. B} \textbf{\bibinfo{volume}{34}},
  \bibinfo{pages}{979} (\bibinfo{year}{1986}).

\bibitem[{\citenamefont{Wunsch et~al.}(2006)\citenamefont{Wunsch, Stauber,
  Sols, and Guinea}}]{Wunsch-NJP06}
\bibinfo{author}{\bibfnamefont{B.}~\bibnamefont{Wunsch}},
  \bibinfo{author}{\bibfnamefont{T.}~\bibnamefont{Stauber}},
  \bibinfo{author}{\bibfnamefont{F.}~\bibnamefont{Sols}}, \bibnamefont{and}
  \bibinfo{author}{\bibfnamefont{F.}~\bibnamefont{Guinea}},
  \bibinfo{journal}{New Journal of Physics} \textbf{\bibinfo{volume}{8}},
  \bibinfo{pages}{318} (\bibinfo{year}{2006}).

\bibitem[{\citenamefont{Hwang and Sarma}(2007)}]{Hwang-PRB07}
\bibinfo{author}{\bibfnamefont{E.}~\bibnamefont{Hwang}} \bibnamefont{and}
  \bibinfo{author}{\bibfnamefont{S.~D.} \bibnamefont{Sarma}},
  \bibinfo{journal}{Phys. Rev. B} \textbf{\bibinfo{volume}{75}},
  \bibinfo{pages}{205418} (\bibinfo{year}{2007}).

\bibitem[{\citenamefont{Stauber}(2014)}]{Stauber-JPCM14}
\bibinfo{author}{\bibfnamefont{T.}~\bibnamefont{Stauber}}, \bibinfo{journal}{J.
  Phys.: Condens. Matter} \textbf{\bibinfo{volume}{26}},
  \bibinfo{pages}{123201} (\bibinfo{year}{2014}).

\bibitem[{\citenamefont{Rold\'an and Brey}(2013)}]{Roldan-PRB13}
\bibinfo{author}{\bibfnamefont{R.}~\bibnamefont{Rold\'an}} \bibnamefont{and}
  \bibinfo{author}{\bibfnamefont{L.}~\bibnamefont{Brey}},
  \bibinfo{journal}{Phys. Rev. B} \textbf{\bibinfo{volume}{88}},
  \bibinfo{pages}{115420} (\bibinfo{year}{2013}).

\bibitem[{\citenamefont{Tabert and Nicol}(2014)}]{Tabert-PRB14}
\bibinfo{author}{\bibfnamefont{C.~J.} \bibnamefont{Tabert}} \bibnamefont{and}
  \bibinfo{author}{\bibfnamefont{E.~J.} \bibnamefont{Nicol}},
  \bibinfo{journal}{Phys. Rev. B} \textbf{\bibinfo{volume}{89}},
  \bibinfo{pages}{195410} (\bibinfo{year}{2014}).

\bibitem[{\citenamefont{Scholz et~al.}(2013)\citenamefont{Scholz, Stauber, and
  Schliemann}}]{Scholz-PRB13}
\bibinfo{author}{\bibfnamefont{A.}~\bibnamefont{Scholz}},
  \bibinfo{author}{\bibfnamefont{T.}~\bibnamefont{Stauber}}, \bibnamefont{and}
  \bibinfo{author}{\bibfnamefont{J.}~\bibnamefont{Schliemann}},
  \bibinfo{journal}{Phys. Rev. B} \textbf{\bibinfo{volume}{88}},
  \bibinfo{pages}{035135} (\bibinfo{year}{2013}).

\bibitem[{\citenamefont{Groenewald et~al.}(2016)\citenamefont{Groenewald,
  R{\"o}sner, Sch{\"o}nhoff, Haas, and Wehling}}]{Groenewald-PRB16}
\bibinfo{author}{\bibfnamefont{R.}~\bibnamefont{Groenewald}},
  \bibinfo{author}{\bibfnamefont{M.}~\bibnamefont{R{\"o}sner}},
  \bibinfo{author}{\bibfnamefont{G.}~\bibnamefont{Sch{\"o}nhoff}},
  \bibinfo{author}{\bibfnamefont{S.}~\bibnamefont{Haas}}, \bibnamefont{and}
  \bibinfo{author}{\bibfnamefont{T.}~\bibnamefont{Wehling}},
  \bibinfo{journal}{Phys. Rev. B} \textbf{\bibinfo{volume}{93}},
  \bibinfo{pages}{205145} (\bibinfo{year}{2016}).

\bibitem[{\citenamefont{Low et~al.}(2014)\citenamefont{Low, Rold\'an, Wang,
  Xia, Avouris, Moreno, and Guinea}}]{Low-PRL14}
\bibinfo{author}{\bibfnamefont{T.}~\bibnamefont{Low}},
  \bibinfo{author}{\bibfnamefont{R.}~\bibnamefont{Rold\'an}},
  \bibinfo{author}{\bibfnamefont{H.}~\bibnamefont{Wang}},
  \bibinfo{author}{\bibfnamefont{F.}~\bibnamefont{Xia}},
  \bibinfo{author}{\bibfnamefont{P.}~\bibnamefont{Avouris}},
  \bibinfo{author}{\bibfnamefont{L.~M.} \bibnamefont{Moreno}},
  \bibnamefont{and} \bibinfo{author}{\bibfnamefont{F.}~\bibnamefont{Guinea}},
  \bibinfo{journal}{Phys. Rev. Lett.} \textbf{\bibinfo{volume}{113}},
  \bibinfo{pages}{106802} (\bibinfo{year}{2014}).

\bibitem[{\citenamefont{Ando et~al.}(1982)\citenamefont{Ando, Fowler, and
  Stern}}]{AFS82}
\bibinfo{author}{\bibfnamefont{T.}~\bibnamefont{Ando}},
  \bibinfo{author}{\bibfnamefont{A.~B.} \bibnamefont{Fowler}},
  \bibnamefont{and} \bibinfo{author}{\bibfnamefont{F.}~\bibnamefont{Stern}},
  \bibinfo{journal}{Rev. Mod. Phys.} \textbf{\bibinfo{volume}{54}},
  \bibinfo{pages}{437} (\bibinfo{year}{1982}).

\bibitem[{\citenamefont{Giuliani and Vignale}(2005)}]{GV05}
\bibinfo{author}{\bibfnamefont{G.~F.} \bibnamefont{Giuliani}} \bibnamefont{and}
  \bibinfo{author}{\bibfnamefont{G.}~\bibnamefont{Vignale}},
  \emph{\bibinfo{title}{Quatum Theory of the Electron Liquid}}
  (\bibinfo{publisher}{CUP, Cambridge}, \bibinfo{year}{2005}).

\bibitem[{\citenamefont{Rodin and Castro~Neto}(2015)}]{Rodin-PRB15}
\bibinfo{author}{\bibfnamefont{A.~S.} \bibnamefont{Rodin}} \bibnamefont{and}
  \bibinfo{author}{\bibfnamefont{A.~H.} \bibnamefont{Castro~Neto}},
  \bibinfo{journal}{Phys. Rev. B} \textbf{\bibinfo{volume}{91}},
  \bibinfo{pages}{075422} (\bibinfo{year}{2015}).

\bibitem[{\citenamefont{Saberi-Pouya et~al.}(2016)\citenamefont{Saberi-Pouya,
  Vazifehshenas, Farmanbar, and Salavati-fard}}]{Pouya-JPCM16}
\bibinfo{author}{\bibfnamefont{S.}~\bibnamefont{Saberi-Pouya}},
  \bibinfo{author}{\bibfnamefont{T.}~\bibnamefont{Vazifehshenas}},
  \bibinfo{author}{\bibfnamefont{M.}~\bibnamefont{Farmanbar}},
  \bibnamefont{and}
  \bibinfo{author}{\bibfnamefont{T.}~\bibnamefont{Salavati-fard}},
  \bibinfo{journal}{Journal of Physics: Condensed Matter}
  \textbf{\bibinfo{volume}{28}}, \bibinfo{pages}{285301}
  (\bibinfo{year}{2016}).

\bibitem[{\citenamefont{Pyatkovskiy and Chakraborty}(2016)}]{Pyatkovskiy-PRB16}
\bibinfo{author}{\bibfnamefont{P.~K.} \bibnamefont{Pyatkovskiy}}
  \bibnamefont{and}
  \bibinfo{author}{\bibfnamefont{T.}~\bibnamefont{Chakraborty}},
  \bibinfo{journal}{Phys. Rev. B} \textbf{\bibinfo{volume}{93}},
  \bibinfo{pages}{085145} (\bibinfo{year}{2016}).

\bibitem[{\citenamefont{Jin et~al.}(2015{\natexlab{b}})\citenamefont{Jin,
  Rold\'an, Katsnelson, and Yuan}}]{Jin-PRB15}
\bibinfo{author}{\bibfnamefont{F.}~\bibnamefont{Jin}},
  \bibinfo{author}{\bibfnamefont{R.}~\bibnamefont{Rold\'an}},
  \bibinfo{author}{\bibfnamefont{M.~I.} \bibnamefont{Katsnelson}},
  \bibnamefont{and} \bibinfo{author}{\bibfnamefont{S.}~\bibnamefont{Yuan}},
  \bibinfo{journal}{Phys. Rev. B} \textbf{\bibinfo{volume}{92}},
  \bibinfo{pages}{115440} (\bibinfo{year}{2015}{\natexlab{b}}).

\bibitem[{\citenamefont{Yuan et~al.}(2015)\citenamefont{Yuan, Rudenko, and
  Katsnelson}}]{Yuan_PRB_2015}
\bibinfo{author}{\bibfnamefont{S.}~\bibnamefont{Yuan}},
  \bibinfo{author}{\bibfnamefont{A.~N.} \bibnamefont{Rudenko}},
  \bibnamefont{and} \bibinfo{author}{\bibfnamefont{M.~I.}
  \bibnamefont{Katsnelson}}, \bibinfo{journal}{Phys. Rev. B}
  \textbf{\bibinfo{volume}{91}}, \bibinfo{pages}{115436}
  (\bibinfo{year}{2015}).

\bibitem[{\citenamefont{Island et~al.}(2016)\citenamefont{Island, Biele,
  Barawi, Clamagirand, Ares, S{\'a}nchez, van~der Zant, Ferrer, D'Agosta, and
  Castellanos-Gomez}}]{Island_SR_2016}
\bibinfo{author}{\bibfnamefont{J.~O.} \bibnamefont{Island}},
  \bibinfo{author}{\bibfnamefont{R.}~\bibnamefont{Biele}},
  \bibinfo{author}{\bibfnamefont{M.}~\bibnamefont{Barawi}},
  \bibinfo{author}{\bibfnamefont{J.~M.} \bibnamefont{Clamagirand}},
  \bibinfo{author}{\bibfnamefont{J.~R.} \bibnamefont{Ares}},
  \bibinfo{author}{\bibfnamefont{C.}~\bibnamefont{S{\'a}nchez}},
  \bibinfo{author}{\bibfnamefont{H.~S.} \bibnamefont{van~der Zant}},
  \bibinfo{author}{\bibfnamefont{I.~J.} \bibnamefont{Ferrer}},
  \bibinfo{author}{\bibfnamefont{R.}~\bibnamefont{D'Agosta}}, \bibnamefont{and}
  \bibinfo{author}{\bibfnamefont{A.}~\bibnamefont{Castellanos-Gomez}},
  \bibinfo{journal}{Sci. Rep.} \textbf{\bibinfo{volume}{6}},
  \bibinfo{pages}{22214} (\bibinfo{year}{2016}).

\bibitem[{\citenamefont{Xia et~al.}(2014)\citenamefont{Xia, Wang, and
  Jia}}]{Xia_NC_2014}
\bibinfo{author}{\bibfnamefont{F.}~\bibnamefont{Xia}},
  \bibinfo{author}{\bibfnamefont{H.}~\bibnamefont{Wang}}, \bibnamefont{and}
  \bibinfo{author}{\bibfnamefont{Y.}~\bibnamefont{Jia}}, \bibinfo{journal}{Nat.
  Comm.} \textbf{\bibinfo{volume}{5}}, \bibinfo{pages}{4458}
  (\bibinfo{year}{2014}).

\bibitem[{\citenamefont{Quereda et~al.}(2016)\citenamefont{Quereda, San-Jose,
  Parente, Vaquero-Garzon, Molina-Mendoza, Agra{\"\i}t, Rubio-Bollinger,
  Guinea, Rold{\'a}n, and Castellanos-Gomez}}]{Quereda_NL_2016}
\bibinfo{author}{\bibfnamefont{J.}~\bibnamefont{Quereda}},
  \bibinfo{author}{\bibfnamefont{P.}~\bibnamefont{San-Jose}},
  \bibinfo{author}{\bibfnamefont{V.}~\bibnamefont{Parente}},
  \bibinfo{author}{\bibfnamefont{L.}~\bibnamefont{Vaquero-Garzon}},
  \bibinfo{author}{\bibfnamefont{A.~J.} \bibnamefont{Molina-Mendoza}},
  \bibinfo{author}{\bibfnamefont{N.}~\bibnamefont{Agra{\"\i}t}},
  \bibinfo{author}{\bibfnamefont{G.}~\bibnamefont{Rubio-Bollinger}},
  \bibinfo{author}{\bibfnamefont{F.}~\bibnamefont{Guinea}},
  \bibinfo{author}{\bibfnamefont{R.}~\bibnamefont{Rold{\'a}n}},
  \bibnamefont{and}
  \bibinfo{author}{\bibfnamefont{A.}~\bibnamefont{Castellanos-Gomez}},
  \bibinfo{journal}{Nano Lett.} \textbf{\bibinfo{volume}{16}},
  \bibinfo{pages}{2931} (\bibinfo{year}{2016}).

\end{thebibliography}

%
%

\end{document}